\documentclass[12pt]{article}
\newcommand{\ignore}[1]{}
\pdfoutput=1

\usepackage{cite}
\usepackage{bm}
\usepackage{placeins}

\renewcommand{\>}{\rangle}
\newcommand{\<}{\langle}
\renewcommand{\Im}{{\rm Im \, }}

\usepackage[latin9]{inputenc}
\usepackage{geometry}
\usepackage{verbatim}
\usepackage{prettyref}
\usepackage{mathdots}
\usepackage{amsmath,mathtools}
\usepackage[normalem]{ulem}
\usepackage{amssymb}
\usepackage{graphicx}
\graphicspath{{fig/}}
\usepackage{setspace}
\usepackage{esint}
\usepackage{bbold}
\usepackage{dsfont}
\usepackage{pdfpages}
\usepackage{caption}
\usepackage{subcaption}

\usepackage{color}
\definecolor{darkgreen}{rgb}{0,0.5,0}
\definecolor{darkblue}{rgb}{0,0,0.6}
\definecolor{purple}{rgb}{0.4,.2,0.7}
\definecolor{awesome}{rgb}{1.0, 0.13, 0.32}

\numberwithin{equation}{section}

\newcommand\be{\begin{equation}}
\newcommand\ee{\end{equation}}
\newcommand{\bea}{\begin{eqnarray}}
\newcommand{\eea}{\end{eqnarray}}
\newcommand\nn{\nonumber \\}

\DeclareMathOperator{\arctanh}{arctanh}

\DeclareMathOperator{\arccosh}{arccosh}
\DeclareMathOperator{\Li}{Li}

\usepackage[colorlinks=true,citecolor=darkgreen,linkcolor=purple,urlcolor=purple]{hyperref}

\begin{document}

\vskip15mm

\begin{center}

{\Large \textsc{Lattice Setup for Quantum Field Theory in AdS$_2$ }}

\vskip15mm

Richard C. Brower$^{1,2}$, Cameron V. Cogburn$^1$, A. Liam Fitzpatrick$^1$, \\ Dean Howarth$^1$, Chung-I Tan$^3$

\vskip5mm

\it{$^1$Department of Physics, Boston University, Boston, MA 02215-2521, USA}\\
\it{$^2$Center for Computational Science, Boston University, Boston, MA 02215, USA}\\
\it{$^3$Brown Theoretical Physics Center and Department of Physics, Brown University, Providence, RI 02912-1843, USA}

\vskip5mm

\tt{brower@bu.edu, cogburn@bu.edu, fitzpatr@bu.edu, howarth@bu.edu, chung-i\_tan@brown.edu}

\end{center}

\vskip15mm

\begin{abstract}
Holographic Conformal Field Theories (CFTs) are usually studied in a limit where the gravity description is weakly coupled. By contrast, lattice quantum field theory can be used as a tool for doing computations in a wider class of holographic CFTs where gravity remains weak but nongravitational interactions {\it in AdS} become strong.  We take preliminary steps for studying such theories on the lattice by constructing the discretized theory of a scalar field in AdS$_2$ and investigating its approach to the continuum limit in the free and perturbative regimes.  Our main focus is on finite sub-lattices of maximally symmetric tilings of hyperbolic space.  Up to boundary effects, these tilings preserve the triangle group as a large discrete subgroup of AdS$_2$, but have a minimum lattice spacing that is comparable to the radius of curvature of the underlying spacetime.  We quantify the effects of the lattice spacing as well as the boundary effects, and find that they can be accurately modeled by modifications within the framework of the continuum limit description.  We also show how to do refinements of the lattice that shrink the lattice spacing at the cost of breaking the triangle group symmetry of the maximally symmetric tilings.  
\end{abstract}

\pagebreak

\pagestyle{plain}
 
\setcounter{tocdepth}{2}

\setlength{\parskip}{.07in}
\tableofcontents
\setlength{\parskip}{.2in}
\newpage
\section{Introduction}

Our expectations about what kinds of behavior are possible in physical systems are often strongly affected by the tractable examples we have at our disposal.  Conformal Field Theories (CFTs) suffer from the disadvantage that finding examples to work with can be difficult since one has to make sure that all $\beta$ functions vanish.  Consequently, known classes of CFTs tend to be fairly rigid in that they have at most a small number of discrete  parameters.  

By contrast, one of the strengths of AdS/CFT is that the conformal symmetry of the field theory dual is  built into the spacetime isometries of the AdS description for any values of the bulk parameters, so the boundary theory is automatically conformal.  If one is content to work with effective theories in AdS, one can thereby easily scan over large classes of CFTs with many continuously tune-able parameters. These classes can be a fantastic source of concrete models for many types of strongly coupled physics, and in many cases have eventually led to a deeper understanding of the behavior in the field theory which can after the fact be formulated without reference to a bulk dual.   However, these classes of bulk models are still rather special because in order to be calculationally tractable, the bulk theory usually must be taken to be perturbative.\footnote{See \cite{Carmi:2018qzm} for interesting work on bulk theories that are calculable due to large $N$ {\it in the bulk} rather than due to weak coupling.  See also \cite{Aharony:2012jf,Aharony:2010ay,Aharony:2015zea}.}  
We will refer to such theories as ``Large $N$'' CFTs, where $N$ is not necessarily related to the central charge of the CFT but rather refers to the fact that all correlators are Gaussian to leading order in a small expansion parameter ``$1/N$''. 

To go beyond the class of weakly coupled bulk theories, one would like to be able to do strong coupling calculations in AdS.  The best-developed numeric tool for QFT at strong coupling is lattice field theory, and by applying it to strongly coupled QFTs {\it in AdS} one can potentially learn about qualitatively new kinds of CFTs.  An initial objection might be that it is not clear how to include gravity in a lattice calculation, and therefore the CFT duals would not contain a stress tensor.  However, we can still study the limit where gravitational interactions in AdS are weak, and so we can neglect them (or potentially include them perturbatively), whereas some of the nongravitational interactions are strong.  Such theories will still be ``small $N$'' in that correlators of generic operators will not be approximately Gaussian.

The goal of this paper is to  build the basic lattice
scaffolding to do these non-perturbative calculations in AdS.  We will restrict our attention to the case of AdS$_2$, though in principle it should be possible to work in any number of dimensions. The
first step in constructing a lattice field theory is to choose a lattice that respects as well as possible the isometries of the
target manifold and to understand the UV and IR cut-off effects, which are referred to
as lattice spacing and finite volume errors, respectively. Both of these
are more challenging and novel in hyperbolic space than in flat space.

On two-dimensional manifolds of constant curvature, the triangle group provides an elegant approach to maintaining a maximal discrete subgroup
of the AdS$_2$ isometries. The approach is essentially to choose a fundamental
triangle and generate the lattice by reflection on the edges. 
For example a familiar case in flat space  is that starting with an equilateral triangle, one generates
the infinite triangulated lattice.  
In negative curvature hyperbolic space, there are an infinite number of such tessellations, 
some of which are illustrated   in the classic Escher drawings. These hyperbolic graphs play a role in tensor networks~\cite{Evenbly:2017hyg, Jahn:2017tls, Boyle:2018uiv}  and quantum error correction codes~\cite{Swingle:2009bg, Almheiri:2014lwa}  in the context of AdS/CFT,  and we present a sketch of the
triangle group algebra  with the hope of  relevance beyond the
present context.

The main difference with tilings of flat space is that
in hyperbolic space, there is a minimum possible size of the triangles in units of the spatial curvature, i.e.,
the lattice spacing cannot be taken arbitrarily small while preserving the full discrete subgroup of spatial symmetries.  
Consequently, in order to use lattice methods to probe distances shorter than the AdS curvature length $\ell$, we must subdivide (``refine'') our maximally symmetric lattice in a way that breaks the discrete symmetries.  Nevertheless, the maximally symmetric lattices are useful for several reasons.  First, they provide a warm-up case where one can check the lattice calculations in a simpler setting.  The large discrete subgroup of the AdS isometries preserved by the tilings means that the full isometries are often `accidental' symmetries of the theory broken only by high dimension operators that scale away quickly at long distances.  Second, if the initial maximally symmetric tiling has triangles that are not much larger than the spatial curvature, then the effect of the curvature within a single triangle is small and further refinements of the tiling are approximately those of flat space.  Finally, most known physical systems have critical exponents that are ${\cal O}(1)$, which translates to their AdS duals having fields with masses $m$ comparable to the AdS radius.  Therefore their Compton wavelength is generally spread out over sizes similar to $\ell$ and perhaps even the tilings without refinement may give good approximations to boundary CFT observables.

This paper is organized as follows.  In Section \ref{sec:tiling}, we review the triangle group that we use to construct maximally symmetric tilings of hyperbolic space, and set up the discretized action for a scalar field theory on this lattice.   In Section \ref{sec:firstpass}, we characterize the behavior of the classical theory on this lattice and compare to analytic results.  Most of the discussion in Section \ref{sec:firstpass} will focus on the free theory, where we compute various propagators and quantify how they differ on the lattice vs in the continuum, but we also compare lattice and continuum results for a tree-level four-point function in $\lambda \phi^4$ theory.  In Sections \ref{sec:finitevolume} and \ref{sec:ref}, we go beyond our maximally symmetric lattices and study how additional corrections can be included in order to approach the continuum limit; Section \ref{sec:finitevolume} focuses on finite volume effects, and Section \ref{sec:ref} shows how to refine the lattices to make the lattice spacing arbitrarily small in units of the radius of curvature of hyperbolic space. Finally,  in Section \ref{sec:future} we conclude with a discussion of future directions.

\FloatBarrier

\section{Tiling the Hyperbolic Disk}
\label{sec:tiling}

Euclidean de Sitter $\mathbb S^d$ and Anti-de Sitter
$\mathbb H^d$ space have constant positive and negative curvature,
respectively.  Although in the continuum they share with flat space
$\mathbb R^d$ the attribute of being a maximally symmetric Riemann manifold,
they pose new difficulties for lattice construction. A lattice realization
necessitates a scheme used to tile the geometry it represents.  In
$\mathbb R^d$ a common lattice choice is hypercubic (square, cubic,
etc.) with a growing discrete subgroup of the isometries: integer
translation on each axis on the torus and hypercubic rotation by
$\pi/2$. In $\mathbb H^d$ all sites are equivalent; there is no ``origin". 
Thus we already see one additional challenge in tiling a space
with constant negative curvature.  In what follows, 
we detail how to tile AdS by constructing a simplicial
lattice for hyperbolic space using the triangle group.

Although the goal
of this section is to construct a simplicial lattice of 
$\mathbb{H}^2$, it is interesting to see this in the context of
the three maximally symmetric manifolds of either positive,
negative, or zero constant curvature.
The  full Euclidean  plane  $\mathbb R^2$ (adding a point at infinity)
is equivalent to the Riemann  sphere up to a Weyl factor: 
\be
ds^2_{\mathbb R^2} = dx^2 +dy^2 \quad \rightarrow \quad ds^2_{\mathbb 
  S^2} = 4 \ell^2 \frac{dz d\bar z}{(1+ |z|^2)^2}  ,
\ee
where  the scalar curvature $K = 1/\ell^2$ and $\ell$ is the radius of
the sphere.
Similarly, hyperbolic space with $K = - 1/\ell^2$  can be represented as the Euclidean upper 
half-plane (UHP) up
to a Weyl factor $1/y^2$, which can then be mapped to the Poincar\'{e} disk
 \be
ds^2_{UHP}  = \ell^2 \frac{dx^2 + dy^2}{y^2}  \quad \rightarrow \quad
ds^2_{\mathbb{H}^2} =  4 \ell^2 \frac{dz d\bar z}{(1-  |z|^2)^2}  
\ee
by the M\"obius transformation $ w = x +iy = - i ( z +1)/(z -1)$ with $|z|
< 1.$ The Poincar\'{e} disk  is  most convenient for our triangulation. 
 The $PSL(2,\mathbb{Z})$  isometries in the UHP are  real M\"obius transformations,
$w \rightarrow w' =  (a w + b)/(c w + b)$, which are easily mapped  to 
Poincar\'{e} disk coordinates. This includes rotations that are
manifest in polar coordinates on the Poincar\'{e}
disk with $z = r e^{i\theta}$. Unlike the sphere, the boundary at $r = 1$
 is at geodesic infinity, which can be seen from the metric $ds^2
= d\rho^2 + \sinh^2(\rho) d\theta$ with the radial geodesic coordinate
$\rho = \log[(1+r)/(1-r)] $.  We will proceed to use the triangle group to tessellate the Poincar\'{e} disk 
with approximately concentric layers around $r = 0$. 

\subsection{The Triangle Group} 
\label{sec:Triangle}
If $p, \,r,$ and $q$ are integers greater than one, the full triangle group
$\Delta(p,r,q)$ is a group that can be realized geometrically as a
sequence of reflections along the sides of a triangle with angles
$(\pi/p, \, \pi/r, \, \pi/q)$. The {\em proper} rotation
triangle group, $D(p,r,q)$, is  generated by two
elements $S,\,T$  satisfying $S^p = T^q = U^r = 1,$ where $U = ST$. The target space is
tessellated by acting on an initial triangle with angles $(2\pi/p, \,
2\pi/r, \, 2 \pi/ q)$ in each vertex.   The full {\em improper}
triangle  group  includes a $\mathbb{Z}_2$ factor for  reflections $a, b, c$ on the edges,
 with $a^2 = b^2 = c^2 =1$. Two reflections along the sides joining at a vertex give a
rotation around that vertex:  $S = ab, \, T = bc, \, U = ca$. Either way the result is a
uniform triangulation of a manifold with constant  positive, zero or
negative curvature dictated by the sum rule: 
\be
 \frac{\pi}{p} + \frac{\pi}{r}  + \frac{\pi}{q} \quad \left\{
                \begin{array}{ll}
                  > \pi \quad \mbox{ positive curvature} \\
                  = \pi \quad \mbox{ zero curvature} \\
                  < \pi \quad \mbox{ negative   curvature} 
                \end{array}
              \right.
 \ee
The group is finite for positive curvature and infinite for flat and
hyperbolic space~\cite{Coxeter}. 

As a familiar example start with flat space where $D(3,3,3)$
generates the triangular lattice. At a more
fundamental level, we can start with the $D(2,3,6)$ group of
90-60-30 triangles and combine 6 of them together to create
an equilateral triangle with 6
sub-triangles. Dropping the midpoint recovers the
$D(3,3,3)$ equilateral tessellation. In the same spirit $D(2,4,4)$ utilizes the
  right triangle for the standard square lattice and the hexagonal
  lattice is the dual of $D(3,3,3)$.  We are most interested
in the special series $D(2,3,q)$, which divides equilateral hyperbolic triangles
into 6 sub-triangles, thereby generalizing the 90-60-30 degree flat case to a triangle with 
90-60-$(30\cdot\frac{6}{q})$ degree interior angles. An important feature of
hyperbolic/spherical triangles is the
introduction of an intrinsic length scale,  $\ell^2= \pm 1/K$ in terms of the scalar
curvature $K$.   This fixes the area  of the hyperbolic and spherical
triangles,
\be
A_\triangle = |\pi - \alpha - \beta - \gamma| \ell^2 \, ,
\label{eq:HypTriangleArea}
\ee
in terms of the interior angles $(\alpha,\beta,\gamma)$ and 
 the intrinsic scale $\ell$.  Whereas in flat space the scaling gives no condition
 on the area, in the spherical case there is
a maximum triangle  area of $2\pi \ell^2$ given by a great circle.
Perhaps a bit surprisingly, in hyperbolic space there is
also a maximum area of $\pi \ell^2$ when all interior angles vanish.

\begin{figure}
\begin{center}
\includegraphics[width=0.5\textwidth, angle=0,scale=0.99]{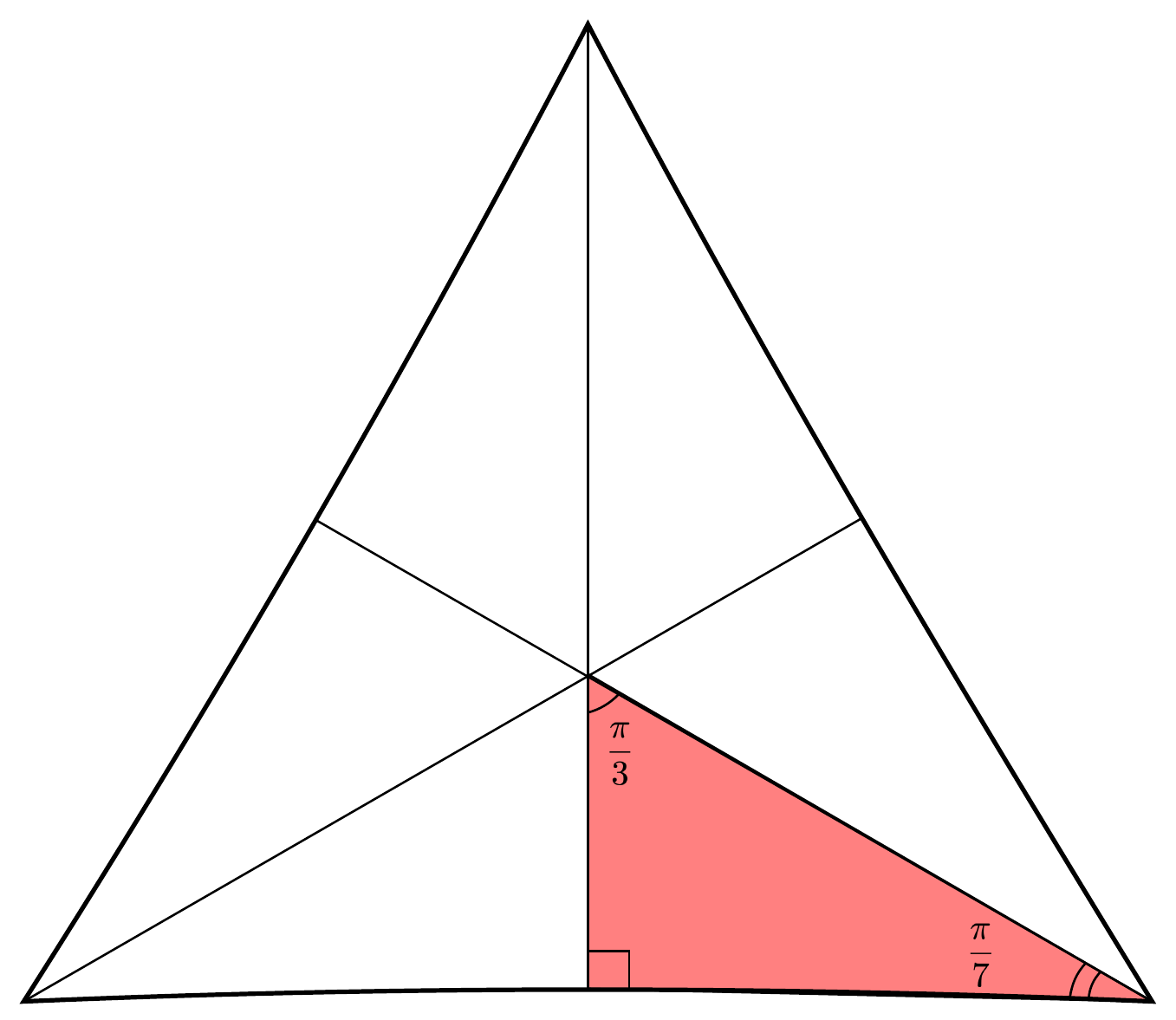}
\includegraphics[width=0.5\textwidth, angle=0,scale=0.99]{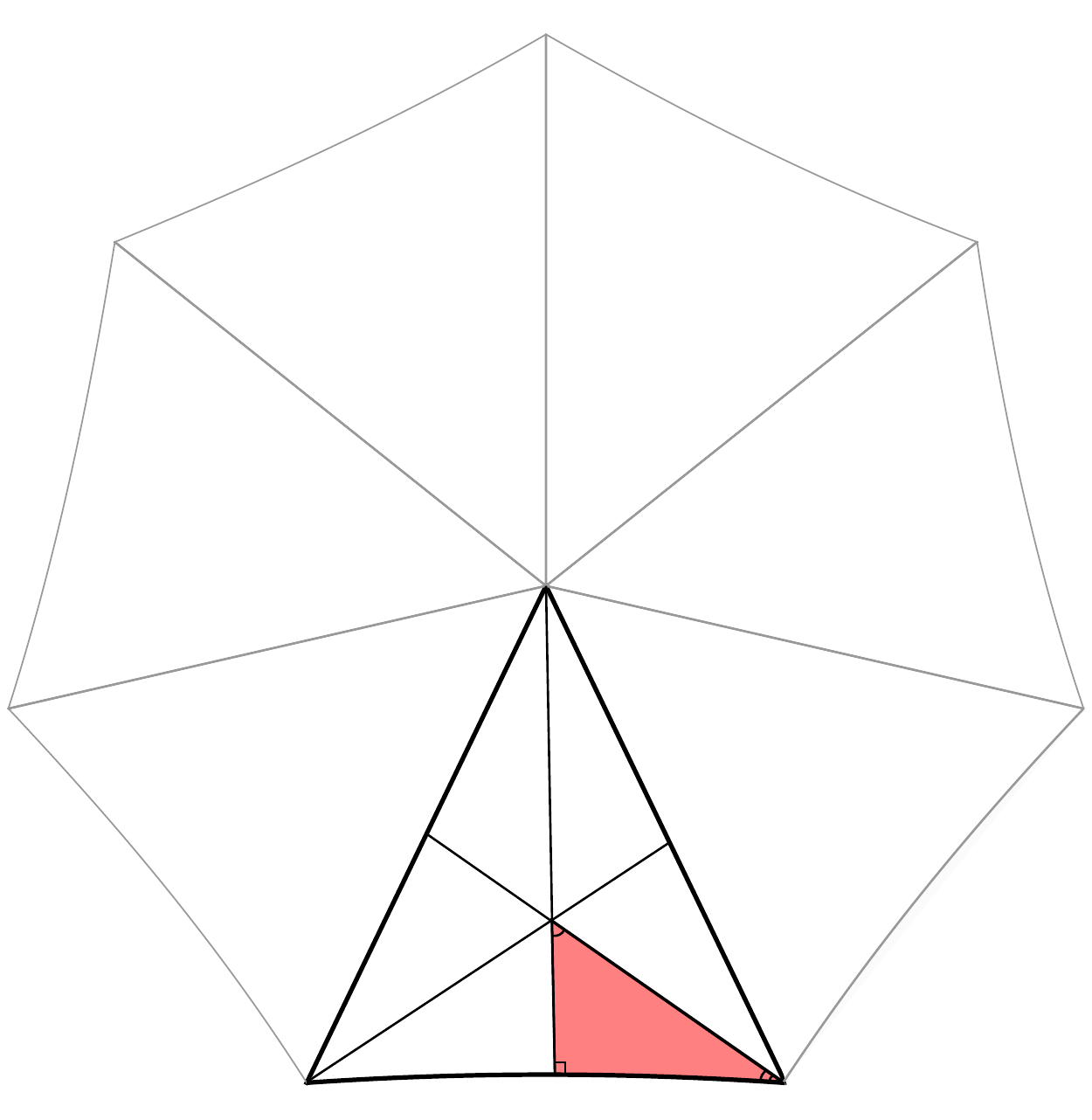}
\caption{We construct the lattice using hyperbolic equilateral triangles built from the $\Delta(2, 3, q)$ triangle group. \textit{Left:} The hyperbolic equilateral triangle ($2\pi/7, \, 2\pi/7, \, 2\pi/7)$ built from the $\Delta(2, 3, 7)$ triangle group. \textit{Right:} The first layer of the lattice constructed by rotating the ($2\pi/7, \, 2\pi/7, \, 2\pi/7)$ equilateral triangle around a point.}
\label{fig:equitri}
\end{center}
\end{figure}

For positive curvature, $q = 3, 4, 5$  produces the finite group
symmetries and tessellations of the \verb! tetrahedron, octahedron! and \verb!icosahedron!, respectively.
The \verb!cube! and the \verb!dodecahedron! are the respective dual polyhedra of the octahedron and icosahedron, thus completing the five Platonic solids.

The situation for $\mathbb H^2$ is more interesting. Even restricting
ourselves to equilateral triangle graphs, starting with the
infinite sequence $D(2,3,q)$ for any $q \ge 7$ gives an 
infinite discrete subgroup\footnote{These groups
have fascinating properties studied exhaustively in 
the mathematics  literature~\cite{Levy,HPSG,Jones}. The full modular group, $PSL(2,\mathbb{Z})$, is reached in the limit
$q\rightarrow \infty$.  In special cases they yield finite
representations of constant negative 
curvature Riemann surfaces - the negative curvature analogue of
Platonic solids.   A famous example is the 168 element tessellation of the genus 3 surface, first introduced  by Felix Klein in 1879, (reprinted in \cite{Levy}), in the study of the Riemann surface associated with the ``Klein quartic", $x^3y + y^3z + z^3 x=0$. For a  historical account of this work, see \cite{Levy,HPSG,Jones,Mumford}.}
of hyperbolic isometries $PSL(2,\mathbb{R})$.  The smallest triangles with
the smallest curvature defect at the vertices is  given by $q = 7$ and will be the
major focus of our lattice construction.

\subsection{Constructing the Lattice}

From the above discussion, the hyperbolic plane can be tiled with
$\Delta(2,3,q)$ where $q$ is any integer with $q \geq 7$. One of the
benefits of a lattice built up from $\Delta(2,3,q)$ is that there is a
$\mathbb{Z}_q$ symmetry. This can be exploited to simplify
calculations and also provides a check for the structure we expect in
certain data. The case $q=7$ is particularly nice as it gives
triangles in AdS$_2$ with the smallest area and therefore the least
curvature (cf.\ Appendix \ref{app:hyptrig} for details). We mention in passing that the case $q=8$ is also
useful in that a geodesic connecting two points will always pass
through lattice points.

\begin{figure}
\begin{center}
\includegraphics[width=0.4\textwidth]{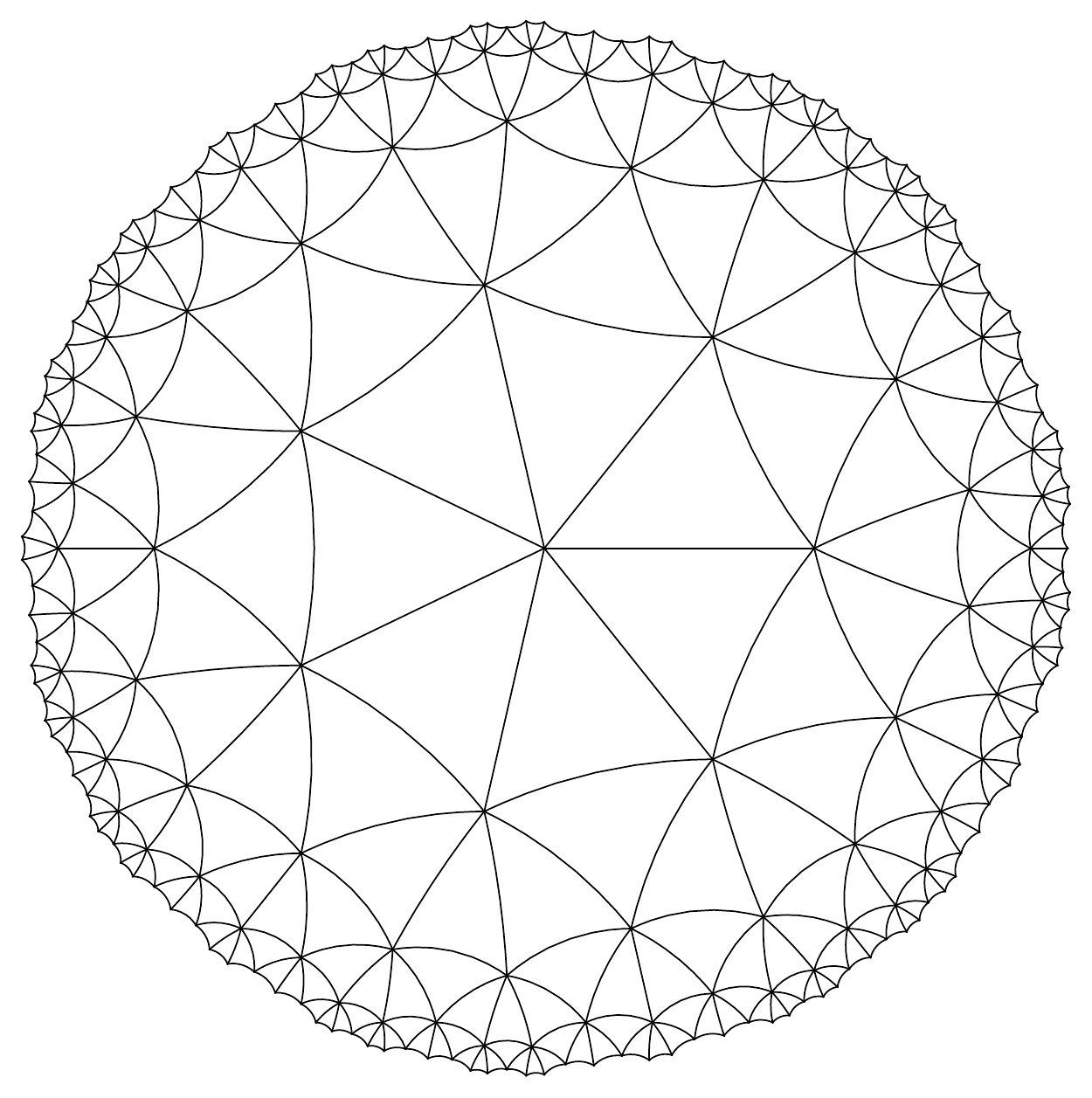}
\includegraphics[width=0.4\textwidth]{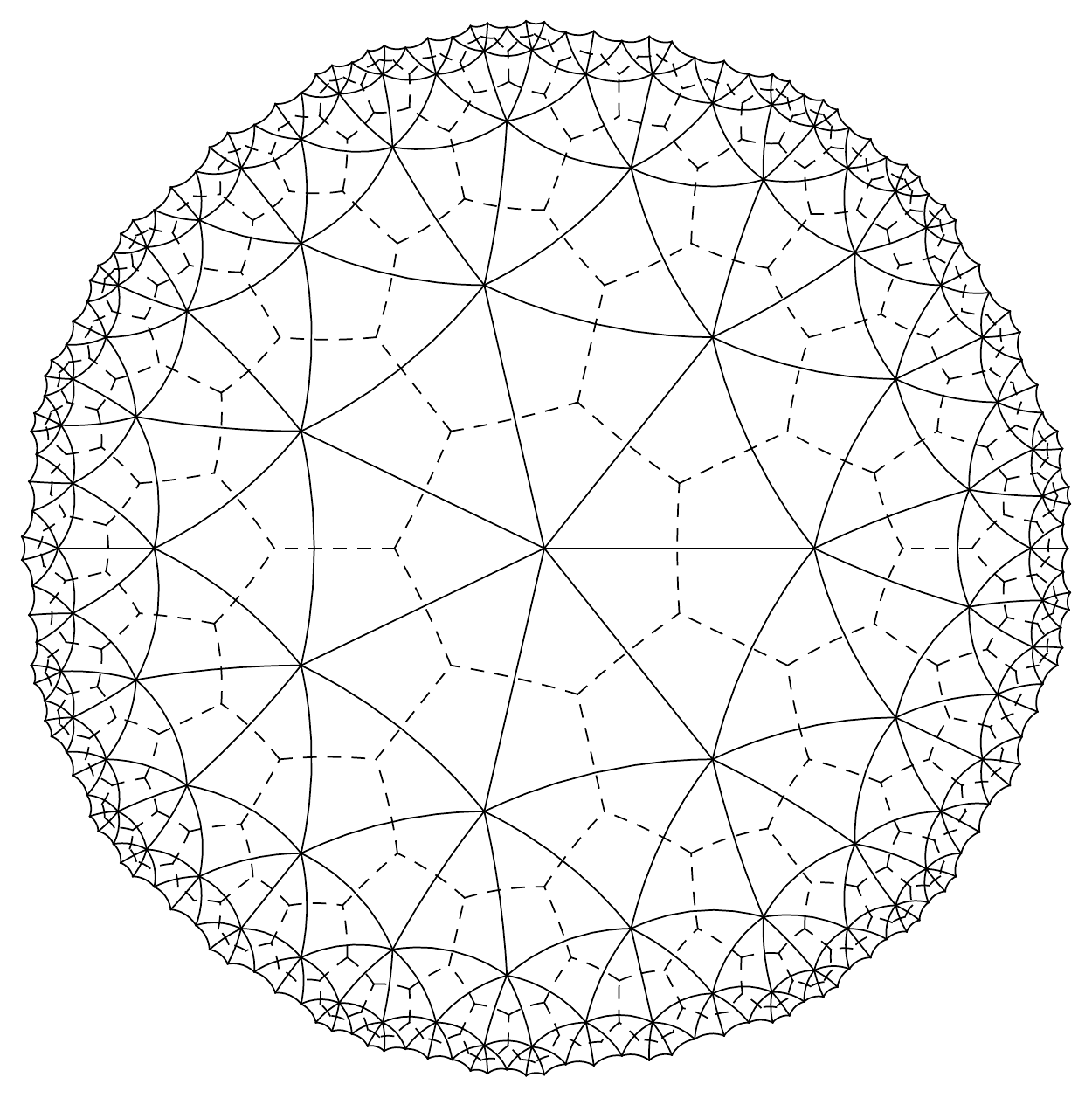}
\caption{\textit{Left:} The $L=4$, $q=7$ lattice on the Poincar\'{e} disk. \textit{Right:} The same lattice with the dual lattice shown in dotted lines. In the infinite $L$ limit we produce a tiling of the entire Poincar\'{e} disk.}
\label{fig:PDtri}
\end{center}
\end{figure}

In practice the lattice is constructed as follows:

\begin{itemize}
\item Choose $q$, which sets the properties of the equilateral triangle such as its area and side length.
\item Put the first lattice point at the origin ($r=0$) of the Poincar\`{e}
  disk. Put the second on the real axis a distance of the side length
  of the equilateral triangle.
\item Rotate by $2\pi/q$ around the origin to add additional triangles
  until the layer is complete.
\item Move out to a point on the next layer. Rotate around this point
  by $2\pi/q$ but laying triangles only along forward links connecting
  to the following layer.
\item Continue until the input layer $L$ is reached.
\end{itemize}

Fig.\ \ref{fig:PDtri} shows a $L=4$, $q=7$ tiling on the Poincar\`{e}
disk and its dual realized from this construction.  It is also
possible to generate the lattice using the action of the triangle
group on the lattice (cf.\ Appendix \ref{app:trigrouplattice}). For
generating large lattices using C-code we use the first method; for
smaller lattices to which we will later add refinements, we use
\verb!Mathematica! utilizing the second method. 
Finally, we note that lattices can also be  generated starting with
the circumcenter of the equilateral triangle at the center ($r=0$). However
with a finite number of layers,  this choice preserves for any q only an exact 3-fold rotational symmetry, 
rather than preserving a q-fold symmetry.

\FloatBarrier
\subsection{Lattice Action for Scalar Fields}
We choose $\phi^4$ theory as a simple example with the action
\be
S = \frac{1}{2} \int d^2 x \sqrt{g} (g^{\mu \nu} \partial_{\mu} \phi \partial_{\nu} \phi + m^2 \phi^2 + \lambda \phi^4)  .
\ee
 Discretizing this action on a 2d  triangulated lattice (or indeed 
any $d$-dimensional  simplicial complex)  takes the generic  form,
\be \label{eq:lattice_action}
S = \frac{1}{2}\sum_{\langle ij \rangle} K_{ij}  (\phi_{i} - \phi_{j})^2 +
\frac{1}{2} \sum_{i} \sqrt{g_i} (m^2 \phi^2_{i} + \lambda \phi^4_{i}) ,
\ee
where the sum is over the sites $i$ and the links $\langle ij \rangle$ on the
graph.  The coefficients $\sqrt{g_i}$ and $K_{ij}$ can be determined by the
finite element method literature (FEM)~\cite{StrangFix200805} as
presented in detail for positive curvature triangulation in
Ref.\cite{Brower:2016vsl,Brower:2018szu}. This method sets the
measure $\sqrt{g_i}$ to the volume at dual sites (e.g.\ the heptagons in Fig.~\ref{fig:PDtri}) and the kinetic
weights $K_{ij}$ to the ratio $S_{ij} /l_{ij}$ of the  dual to the link (the Hodge star of 
the link $l^{*}_{ij})$ divided by length of the link $l_{ij}$.   In fact,
these kinetic weights $K_{ij} $ were anticipated in the pioneering paper
by N. Christ, R. Frieberg and T.D. Lee~\cite{Christ:1982ci} by
enforcing a discrete Gauss' law constraint on a random lattice and
later generalized as a consequence of discrete exterior calculus
(DEC)~\cite{2005math8341D} on a simplicial complex.

\begin{figure}
\begin{center}
\includegraphics[width=0.55\textwidth]{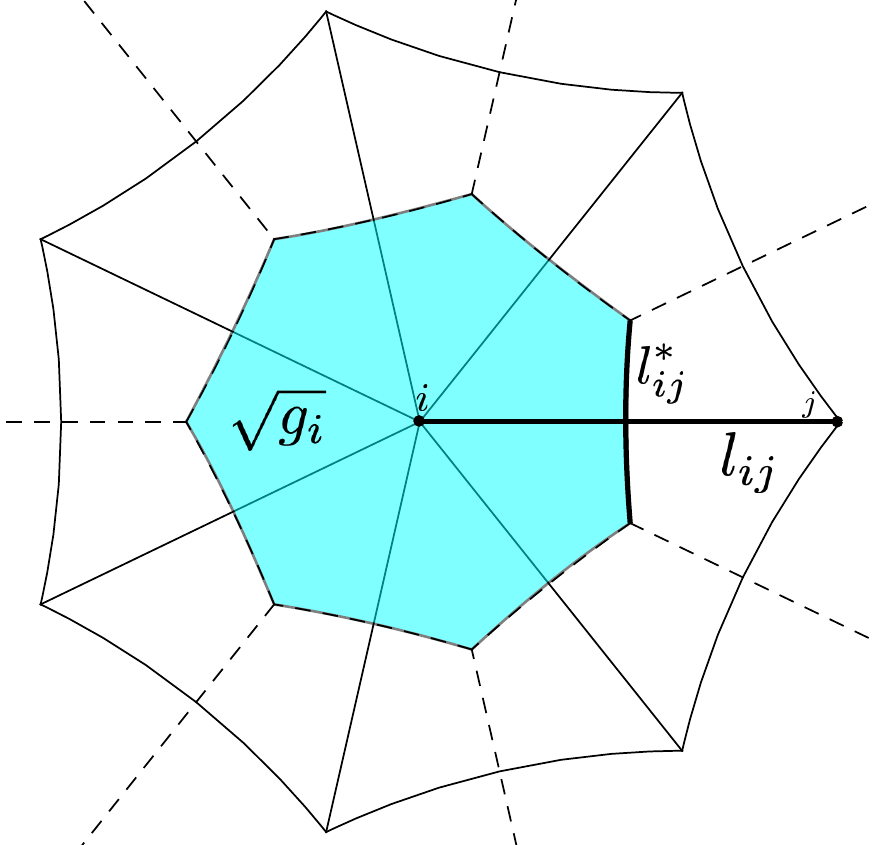}
\caption{The lattice and its dual at a vertex point $i$. The area of the dual heptagon (light blue region) corresponds to $\sqrt{g_{i}}$ and the kinetic weight between $i$ and another point $j$ is given by $K_{ij} = l_{ij} / l^{*}_{ij}$.}
 \label{fig:latanddual}
\end{center}
\end{figure}

For our $\Delta(2,3,q)$ example, illustrated in Fig.~\ref{fig:latanddual} for $q = 7$, the tessellation consists of
uniform equilateral triangles so that both  $\sqrt{g_i}$  and $K_{ij}$ are  constants independent of position. On the hyperbolic disk they are given as
\be 
\sqrt{g_i} =
\frac{q}{3} A_\triangle , \quad  \quad K_{ij} = \frac{4 A_\triangle }{3a^2}  ,
\label{eq:FEMweight}
\ee
in terms of the area $A_\triangle = (\pi - 6 \pi/q) \ell^2$ of the
equilateral triangle  and the lattice spacing $a$, which is the length of each side of the triangle, defined below in (\ref{eq:latticeLength}).  In our numerical work we have adopted the common practice
of setting $K_{ij} = 1$ by an appropriate re-scaling of the field $\phi$. 
 We also have introduced a dimensionless 
 bare mass parameter, 
 \be
 m^2_0 \equiv c_q^2 m^2  , \qquad c_q^2 = \frac{\sqrt{g_i}}{K_{ij}}  = \frac{q a^2}{4}  ,
 \label{eq:m0def}
 \ee 
in terms of  an effective lattice spacing $c_q = q^{1/2}  a/2$, as well as a re-scaled coupling 
 $\lambda_0 =  3 q a^4 \lambda/(16A_\triangle) $. Now in our
 simulations in the triangle group lattice, the action takes the convenient form
\be
S =\frac{1}{2}  \sum_{\<ij\>}
(\phi_i - \phi_j)^2 + \frac{1}{2} \sum_{i}  (m^2_0 \phi^2_{i}
+ \lambda_0 \phi^4_{i})  .
\ee
Restoring the dimensionful form is easily done when we 
compare the simulation to the continuum.

All lattice calculations in flat space have to deal with both lattice
spacing (UV) and finite volume (IR) errors. In flat space, on a toroidal
hypercubic lattice, the lattice spacing $a$ for a finite volume $V = (aL)^d$ has a
single integer ratio $L = V^{1/d}/a$, which is the number of lattice
spacings in a closed cycle. Since  flat space has no intrinsic scale the
lattice spacing  is arbitrary and only defined relative to 
some physical mass scale $m^{-1} = a \xi$. Control over UV and IR
errors requires a window  such that $a \ll m^{-1}
\ll a L$. 

Hyperbolic space (or de Sitter space) is different
because the manifold itself has an intrinsic length scale given by the
curvature. Thus even the infinite lattice has two scales, the UV
cut-off and the curvature, which complicates the problem of approximating a
continuously smooth manifold of constant negative curvature with
objects of a finite size. Additionally, there is no analogue to periodic
boundary conditions to {\em hide} the finite volume boundary. Instead, the boundary 
is an important feature exploited by AdS/CFT, which we discuss later.

The intrinsic continuum scale $\ell$ of the manifold  is provided by
the metric (\ref{eq:hyp_metric}). The area of a  hyperbolic (or
spherical)   triangle $A_\triangle$ is determined by
the deficit angle $|\pi - \alpha - \beta  - \gamma|$. In units of
$\ell^2$ the area is bounded by $A_\triangle/\ell^2 < \pi$.  For our $(2,3,q)$  equilateral
triangulation of the disk the length of the edges $a$ is
given by
\be 
\cosh(a/ 2 \ell )  =  \frac{\cos(\pi/3)}{\sin(\pi/q)}  = \frac{1}{2 \sin(\pi/q)},
\label{eq:latticeLength}
\ee
which gives the minimum values of $ a =1.090550  \ell$ and $A_\triangle=
0.448799  \ell^2$ for $q=7$. Note that as $q \rightarrow \infty$ the
length  diverges logarithmically as $a \cong 2 \ell\log(q) $ but the area
approaches the so-called ideal  hyperbolic triangle area of $A_\triangle=
\pi  \ell^2$. We focus exclusively on the case $q = 7$ that gives the minimum
intrinsic lattice scale.

\subsection{Lattice Simulations}

Our lattice simulation generates a graph by starting with a vertex at $r
= 0$ and building out one layer of triangles at a time to $L$ layers. Each
layer has $n[L]$ vertices, where $n[L]$ satisfies the recursion relation (cf.\ Appendix \ref{app:Recursive})
\be
n[L] = (q - 4) n[L-1]  -n[L-2]  , \quad \quad n[1] = q , \quad \quad
n[0] = 0 .
\label{eq:Recursion}
\ee
In this graph the growth of vertices is exponential in $L$. Additionally, at asymptotically large $L$ the last
layer has  a finite fraction of all the points in our
lattice.  For $q =7$ this fraction is the inverse of the Golden Ratio, $2/(1+\sqrt{5}) \approx 61 \%$ of all points.  For a finite number of layers, our lattice has only a finite volume and fills out the Poincar\'e disk with a
cut-off, i.e. $|z| \le 1- \epsilon$.  We place fields $\phi_i$ on the
 interior vertices, 
labeled by $r_i, \theta_i$, and impose
Dirichlet boundary conditions on a fictitious $(L+1)$-th layer. 
Not all points on this layer are at the same distance from the `origin', since our lattice breaks rotational invariance to a discrete subgroup.  Nevertheless, we can define an approximate cutoff in $1-|z|$ by taking its value averaged over all points on this fictitious layer, where the boundary condition is enforced:
\be
\epsilon_{eff,L} =  {\rm mean}_{i \in  (L+1)\textrm{-th} \textrm{ layer }} \Big( 1- |z_i| \Big) 
\propto  e^{- 0.919 \ell (L+1)}.
\label{eq:EpsScaling}
\ee
This IR cut-off  $\epsilon_{eff,L} \propto e^{ - \tilde c_7 \ell (L+1)}$ can be thought of as an  effective
lattice length $\tilde c_7 = 0.919 $ in the radial direction.  This  numerical estimate, extrapolating from finite $L$, is very  close to the  asymptotic lattice spacing, $a_7 = 0.962$, between layers in (\ref{eq:efflatspacing}).  Also note that $\tilde{c}_7$ 
is approximately two-thirds of the  UV scale  $c_7 = \sqrt{q/4} (a/\ell) = 1.443$ that we introduced above.

\FloatBarrier
\section{Classical Theory on Triangle Group Lattice}
\label{sec:firstpass}

In this section we mostly study the free discretized theory in AdS$_2$, the
action (\ref{eq:lattice_action}) with $\lambda = 0$, with a  focus on the
errors introduced by the UV cut-off and IR finite volume effects. The continuum limit in the free theory is of course
solvable and will allow us to check our methods and analyze
discretization errors, and will give insight into features that are
absent in the flat space formulation.  Moreover, the main quantities
of interest in the free theory are propagators, which we will use in
later sections to do perturbative computations at weak coupling.

\subsection{Green's Function}

We begin by considering the bulk-to-bulk propagator in AdS.  Both the
bulk-to-boundary propagator and the boundary-to-boundary two-point
function in the dual CFT can be extracted from the bulk Green's
function, so in this sense it is a more basic building block than the
others.  There is another practical reason, however, for beginning
with the bulk propagator, namely our finite lattices can never truly
reach infinity and consequently we will have to be more careful than
usual when extracting propagators with points on the boundary.

We begin with a brief review of some well-known results in the continuum limit.  For a given mass-squared $m^2$, the 
analytic bulk Green's function $G_{bb}(X,X')$ between two points $X, X'$ in AdS$_{d+1}$ is the solution to the equation
\be
(- \nabla^2 + m^2) G = \frac{1}{\sqrt{g}} \delta^{d+1}(X-X'), 
\ee
where $\nabla_\mu$ is the covariant derivative $\frac{1}{\sqrt{g}}\delta^{d+1}(X-X')$ defined as the $\delta$ function for  AdS$_{d+1}$, and the Laplace operator is $\nabla^2 =\nabla_\mu \nabla^\mu = \sqrt{g}^{-1} \partial_\mu
\sqrt{g} g^{\mu \nu} \partial_\nu $ acting on scalars. 
The solution is  given by \cite{Burgess:1984ti,Hijano:2015zsa}
\be \label{eq:exactbulkbulk}
G_{bb}(X, X') = G_{bb}(\sigma(X,X')) = e^{-\Delta \sigma(X,X')} {}_2 F_1 \Big(\Delta, \, \frac{d}{2},\, \Delta + 1 - \frac{d}{2}; \, e^{-2\sigma(X,X')} \Big),
\ee
where $m^2 \ell^2 = \Delta(\Delta-d)$ and $\sigma(X,X')$ is the geodesic distance between the points $X$ and $X'$.
For $d=1$ and $\Delta=1$ this reduces to the simple value of 
\be
G_{bb}(\sigma) \stackrel{\Delta=1}{=} \frac{1}{2} \log \coth \frac{\sigma}{2}.
\ee
For the geodesic distance $\sigma$ in various coordinate systems, see Appendix \ref{app:metric}.

For the most part, we will not need the exact form of the propagator
because the lattice spacing is comparable to the AdS radius and
distinct lattice points have a minimum geodesic distance between them.
Above, we saw that this geodesic distance is smallest for $q=7$ where
it is $\sigma_{\rm min}=a = 1.09 \ell$, and so
$e^{-2\sigma} \le 0.11$ for our lattice points.  We can therefore
approximate the hypergeometric function 
$ {}_2 F_1 \Big(\Delta, \, \frac{1}{2},\, \Delta + 1 - \frac{1}{2}; \,
t^2 \Big)$ in the propagator as $\approx 1$. In fact, between $\Delta=1$ and $\Delta=\infty$,
this  propagator interpolates between the simple functions $\frac{1}{t}\tanh^{-1} t$ and
$(1-t^2)^{-\frac{1}{2}}$, and these never deviate from 1 by more than
$\sim 6$ percent over the range $0 < t^2 < 0.11$.  Consequently, we
will often approximate the propagator as simply
\be
G_{bb}(\sigma) \approx e^{-\Delta \sigma}.
\ee
Both the finite lattice spacing as well as the finite number of layers
make the exact bulk Green's function more complicated in the
discretized theory than in the continuum limit, and we will not be
able to write it down in closed form.  However, in the following we
will be able to obtain closed form approximations in various limits.

\subsection{Taylor Expansion of Lattice Action}

The discretized bulk Green's function is simply the inverse $G_{ij}$ of the matrix $A_{ij}$ from the free theory action,
\bea
S &=& \frac{1}{2} \phi_i A_{ij} \phi_j  =\frac{1}{2} [  \sum_{\<ij\>}
(\phi_i - \phi_j)^2 +  \sum_i  m^2_0  \phi_i^2 ] \nn 
&=&
\frac{1}{2} \sum_i  \left[  \frac{1}{2}\sum_{j \textrm{ adjacent to }
    i} (\phi_i - \phi_j)^2  +  m^2_0 \phi_i^2 \right].
\label{eq:LatticeAction2}
\eea
The sum on $\<ij\>$ is over adjacent pairs of points.  Because each
lattice point has $q$ neighbors, the matrix $A_{ij}$ is $q+ m_0^2$ on
the diagonal, $-1$ in off-diagonal entries if $i$ and $j$ are
neighboring points, and 0 otherwise.

We would like to relate the discretized action to the continuum
action. To do this, we can Taylor expand $\phi$ around each point to
leading order in the lattice spacing:
\be
\phi_j-\phi_i \approx r^\mu_{j} \cdot \nabla_\mu \phi(r_i).
\ee
The Taylor expansion uses the covariant derivative $\nabla_\mu$ to account
for the curvature of space, and the magnitude of $\vec{r}_{j}$ should
be the geodesic length between $i$ and $j$. Because there is a
$\mathbb{Z}_q$ subgroup of the rotational symmetry around each point,
all such adjacent distances are the same lattice spacing $a$.
Moreover, it is easy to see that
\be
\sum_{j \textrm{ adjacent to } i} r_{j}^m r_{j}^n  = \frac{q a^2}{2} \delta^{mn} .
\ee
The fact that the RHS must be proportional to $\delta^{mn}$ is due to
the $\mathbb{Z}_q$ symmetry, and the proportionality constant can be
obtained by taking the trace of both sides of the equation.
Therefore, at leading order in $a$, the discretized free action is equivalent to the following continuum action
\be
S \approx   \frac{3 a^2}{4 A_\triangle} \int d^2 x \sqrt{g} \left[  \frac{1}{2}  (\nabla
  \phi)^2  +  \frac{1}{2}  m^2 \phi^2 \right].
\ee
where  we  re-instated
  the original mass parameter $m^2$ using (\ref{eq:FEMweight}).

If the lattice spacing $a$ could be taken arbitrarily small, then we
could simply expand the discretized kinetic term to leading order in
$a$. In our case, the lattice spacing $a$ is fixed by our choice of
triangle group and has a minimum possible value in units of the AdS
radius $\ell$, so cannot be taken arbitrarily small.  Therefore, to
improve the accuracy of the approximation, we may need to keep higher
order terms in the Taylor expansion.  
 The result for $q=7$ is that due to the
$\mathbb{Z}_7$ symmetry, the action expanded to ${\cal O}(a^{12})$ can
be written in terms of powers of the rotationally symmetric
combination $\nabla^2$:
\be
 S  \propto  \int d^2 x \sqrt{g} \left[\frac{1}{2} (m a)^2 \phi^2 -  \frac{1}{2} \phi \left( \sum_{s=1}^6 b_s  (a^2\nabla^2)^s + {\cal O}(a^{14}) \right)\phi  \right] , 
 \label{eq:TaylorAction}
 \ee
where we have found the   coefficients $b_s$ to be
\be
b_{1} = 1 ,~~~b_{2} =  \frac{1}{16},~~~b_{3} = 
\frac{1}{576},~~~b_{4} = -\frac{7}{36,864},~~~b_{5} =
-\frac{7}{30,700},~~~b_{6} =- \frac{77}{88,473,600}.
\ee
Starting at ${\cal O}(a^{14})$ and higher, one finds combinations of derivatives that break rotational invariance but preserve the $\mathbb{Z}_7$.

\begin{figure}
\centering
\includegraphics[width=0.69\textwidth,scale=1.0]{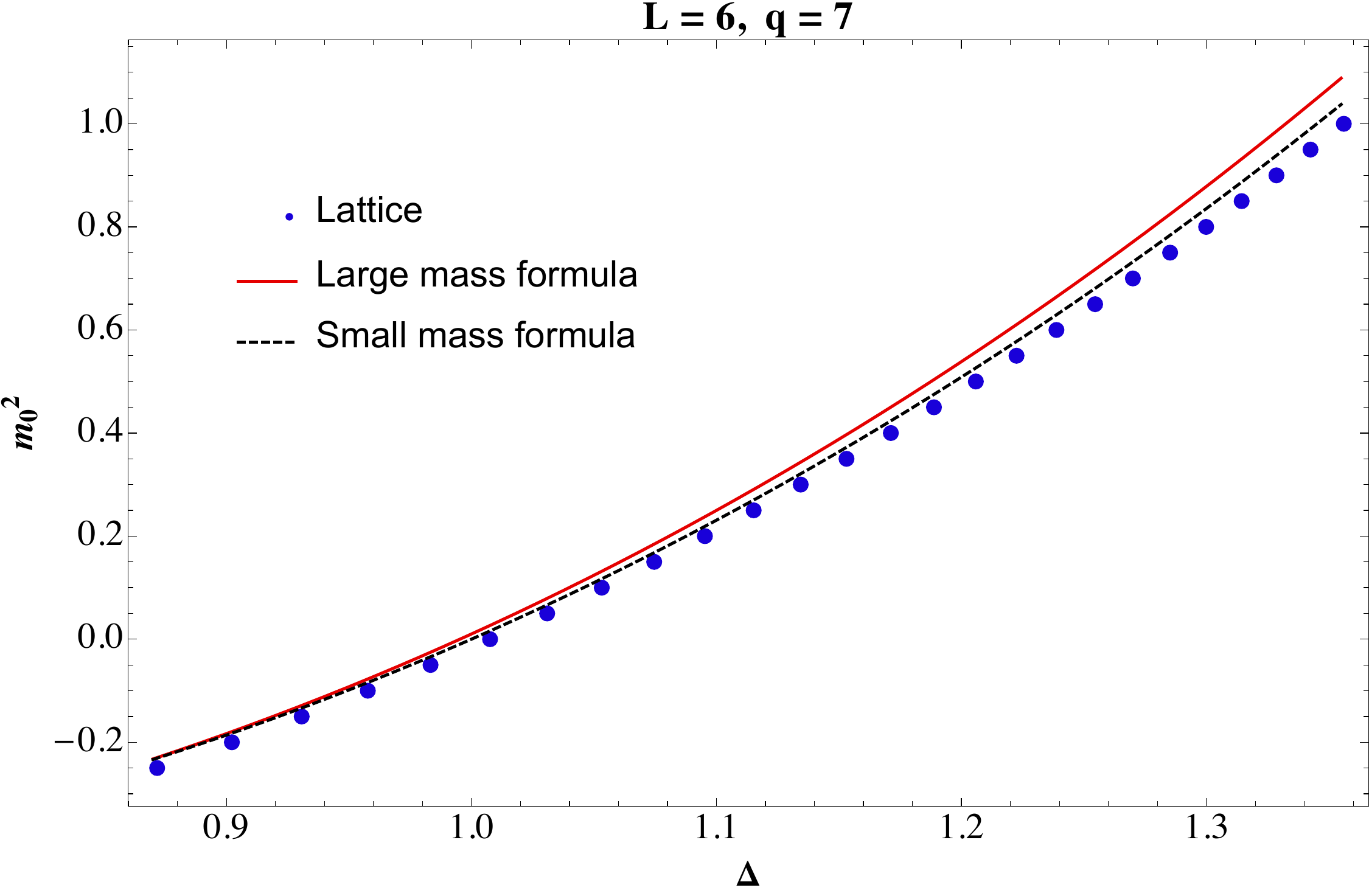}
\caption{A comparison of the mass-squared scaling dimension from the
  lattice data to the expansion formula (\ref{eq:smallmassformula}) and the asymptotic analysis
  formula (\ref{eq:largemassformula}) for small masses. Dots are extracted from lattice data as described in subsection \ref{sec:NumProp}.}
\label{fig:figdeltamass1}
\end{figure}

Now, we can obtain an approximate formula for the discretized
bulk-to-bulk propagator $G_{bb, \, \rm lat}$ that is valid at small
$m^2$.  To do this, recall that (for $\sigma >0$) the continuum
bulk-to-bulk propagator $G_{bb}$ satisfies
\be
\nabla^2 G_{bb} = \frac{1}{\ell^2} \Delta(\Delta-1) G_{bb}.
\ee
If we take $G_{bb}$ with $\Delta$ a free parameter and substitute it
into the equation of motion for our series expanded action
(\ref{eq:TaylorAction}),  we find that it is a solution as long as
\be
m^2 a^2 =  \sum_{s=1}^6 b_s \left( \frac{a^2}{\ell^2} \Delta(\Delta-1)
\right)^s .
\label{eq:smallmassformula}
\ee
Note that $\Delta=1$ remains a solution at $m^2=0$.  By continuity, as
long as $m^2$ is sufficiently small $\Delta$ will be close to 1, and
discarding the ${\cal O}(a^{14})$ terms and higher will be a good
approximation.  In Fig. \ref{fig:figdeltamass1}, we compare the prediction of the analytic equation (\ref{eq:smallmassformula}) with fits to the numerical data, at small mass.  The figure also shows the prediction of another analytic equation valid at large mass, which we derive in the next subsection.

\subsection{Large Mass Effective Model} \label{subsec:LargeMass}

The approximation in the previous subsection for the bulk-to-bulk propagator breaks down when $m^2$ is too large.  Here, we will consider a different approximation that works well for large $m^2$.  
 
Consider the equation of motion for $G_{ij}$ as a function of $j$ with
$i$ fixed, in the limit of large geodesic separation $\sigma$.
With an infinite number of layers, we can take any point $i$ to be the
center of our triangulated lattice without loss of generality. Additionally, near the boundary of AdS
there is a lattice point at every $\theta$ (see Fig. \ref{fig:PDtri}),
but by contrast the lattice spacing in $\rho$ remains the same
regardless of the layer. We can use this fact to obtain an approximate formula for the discretized Green's function $G_{ij}$ at large $\sigma$ by letting $G$ be independent of $\theta$, due to rotational symmetry, and letting  $\rho$ be discrete in the free AdS equation of motion. 

In the continuum the metric can be written (see \ref{app:metric}) as
$ds^2 = d\rho^2 + \sinh^2(\rho) d\theta$. Assuming rotational symmetry the equation of motion is
\be
-\partial^{2}_{\rho} G(\rho) - \coth(\rho)\partial_{\rho}G(\rho) + m^2 G(\rho) = \partial_{\rho}\delta(\rho),
\ee
where $\rho$ is the radial geodesic distance from the source at the origin.

\begin{figure}
\centering
\includegraphics[width=0.69\textwidth,scale=1.0]{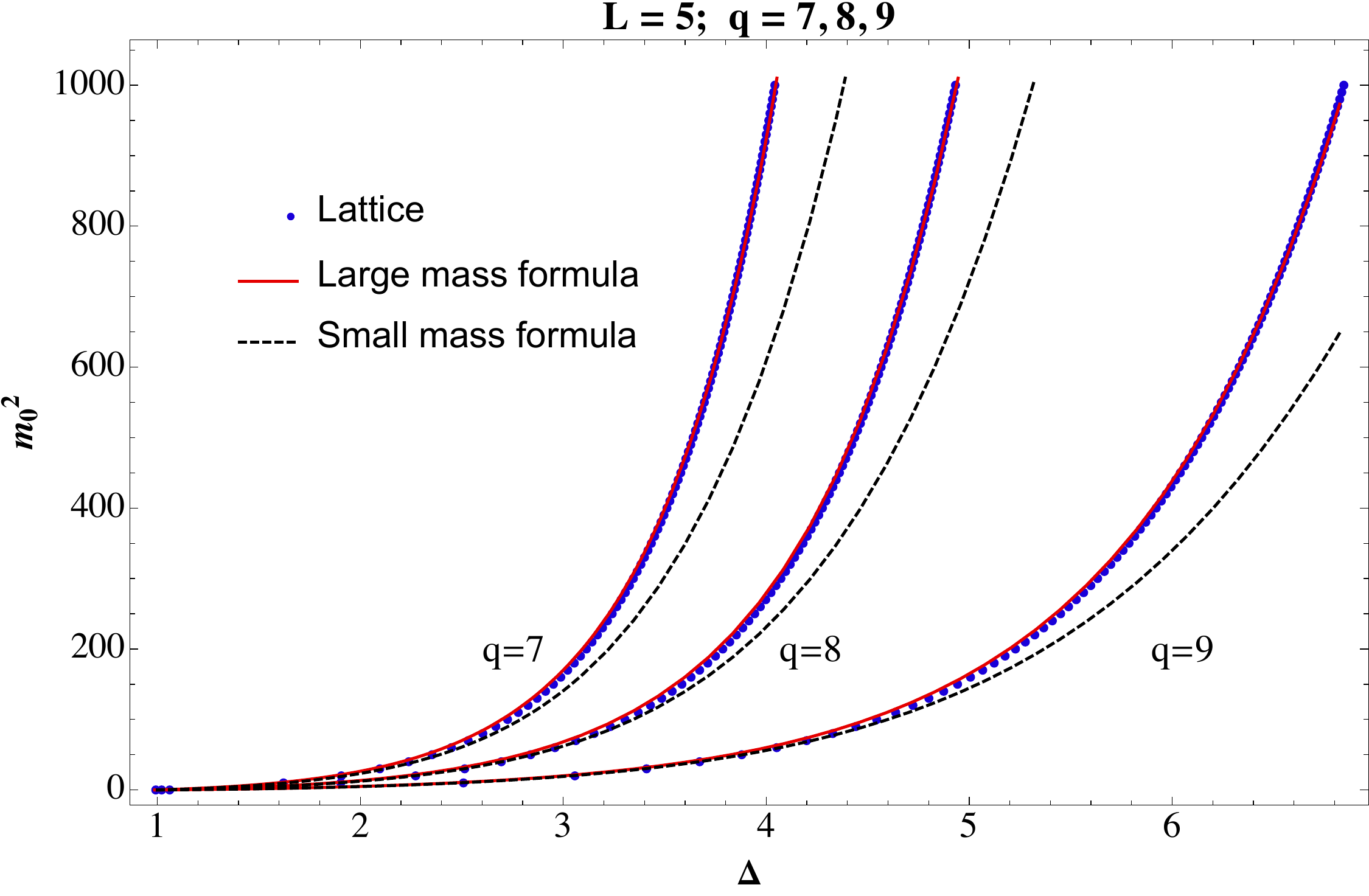}
\caption{Same as Fig. \ref{fig:figdeltamass2} but for large masses.
Dots are extracted from lattice data, black dashed and red solid curves are the expansion formula (\ref{eq:smallmassformula}) and the asymptotic analysis
  formula (\ref{eq:largemassformula}) respectively.}
\label{fig:figdeltamass2}
\end{figure}

At large $\rho$ the Green's function therefore satisfies the
equation,
\be
- \partial_\rho^2G(\rho) - \partial_\rho G(\rho) + m^2  G(\rho)  \cong 0 .
\ee
    Different ways of discretizing the $\partial_\rho$ derivatives lead to different solutions for $G(\rho)$. 
  To see this, let us factor out an overall exponential dependence from $G(\rho)$:
  \be
  G(\rho) = e^{s \rho} g(\rho).
  \ee
  The equation of motion in terms of $g(\rho)$ is
  \be
  s(1+s) g(\rho) + (1+2s) g'(\rho) + g''(\rho) =m^2 \ell^2 g(\rho) .
  \label{eq:EOMwithS}
  \ee
 The LHS of (\ref{eq:EOMwithS}) can be discretized by the following finite difference: 
\be
{\rm LHS} \approx  \frac{(2
   b s+b+2) g(\rho+b )}{2 b^2} +  \left(s^2+s-\frac{2}{b^2}\right) g(\rho )-\frac{(2 b s+b-2) g(\rho -b)}{2 b^2}.
\ee
The resulting finite difference equation, like the continuum equation, has a growing and decaying solution at $\rho \rightarrow \infty$.  The decaying solution behaves at large $\rho$ as $G(\rho) \sim e^{-\Delta \rho}$ for some $\Delta$.  Substituting $g(\rho) = e^{-(\Delta+s)\rho}$ into our discretized equation of motion, we see that the equation is satisfied as long as  $\Delta$ obeys the following equation:
\be
2 (\cosh (b (\Delta +s))-1)-b (2 s+1) \sinh (b (\Delta +s))= b^2 (m^2 \ell^2- s(s+1)) .
\ee  
At large $m^2$, the solution is $\Delta +s \sim b^{-1} \left( \log m^2 \ell^2+ \log\left( \frac{b^2}{1-b(s+\frac{1}{2})}\right)\right) $, whereas in the limit $b\rightarrow 0$ of small lattice spacing, the solution reduces to the usual $m^2 \ell^2  = \Delta(\Delta-1)$.  
The parameters $s$ and $b$ are fudge factors that we use to compensate for the fact that our finite difference equation treats the lattice like a set of regularly spaced points in $\rho$, whereas the true lattice equation involves differences in multiple different directions and is more complicated.  We have found that the choice $s=-1/2$, i.e. 
\be
m^2 \ell^2 =  \left( \frac{4}{b^2}  \sinh^2 \left( b\frac{\Delta-\frac{1}{2}}{2}\right) - \frac{1}{4} \right), \qquad b= \left\{ \begin{array}{cc} 0.96 & 7 \\ 1.34 & 8 \\
1.66 & 9 \end{array} \right.
\label{eq:largemassformula}
\ee
approximates the exact numeric results quite well with the given values of $b$ for $q=7,8,9$, as shown in Fig \ref{fig:figdeltamass2}, where we compare at large masses the analytic approximation (\ref{eq:largemassformula}) with the numeric values of $\Delta$ extracted from the data; in Fig \ref{fig:figdeltamass2}, we also show for comparison the prediction of the small mass analytic expression (\ref{eq:largemassformula}) from the previous subsection. Note that $s=-1/2$ is special in that it respects the $\Delta \rightarrow 1-\Delta$ symmetry of the $m^2$ vs $\Delta$ relation of the continuum theory. Also note that in Appendix \ref{app:Recursive}  we determine analytically  the mean lattice spacings between layers to be $a_q = 0.9624, \,
1.317,\, 1.567$ for  $q = 7,\, 8,\, 9$,  which are nearly identical to the values of $b$ found here by  fitting to the correlator data.

\subsection{Numerical Propagator and Comparison}
\label{sec:NumProp}

Once we have the lattice action in the form (\ref{eq:LatticeAction2}), the numeric computation of the bulk Green's function $G_{ij}$ just requires taking the inverse of $A_{ij}$.  In this subsection, we will inspect the numeric Green's function and see how it compares to the analytic results derived above.

\begin{figure}
\begin{center}
\includegraphics[width=0.5\textwidth, angle=0,scale=0.99]{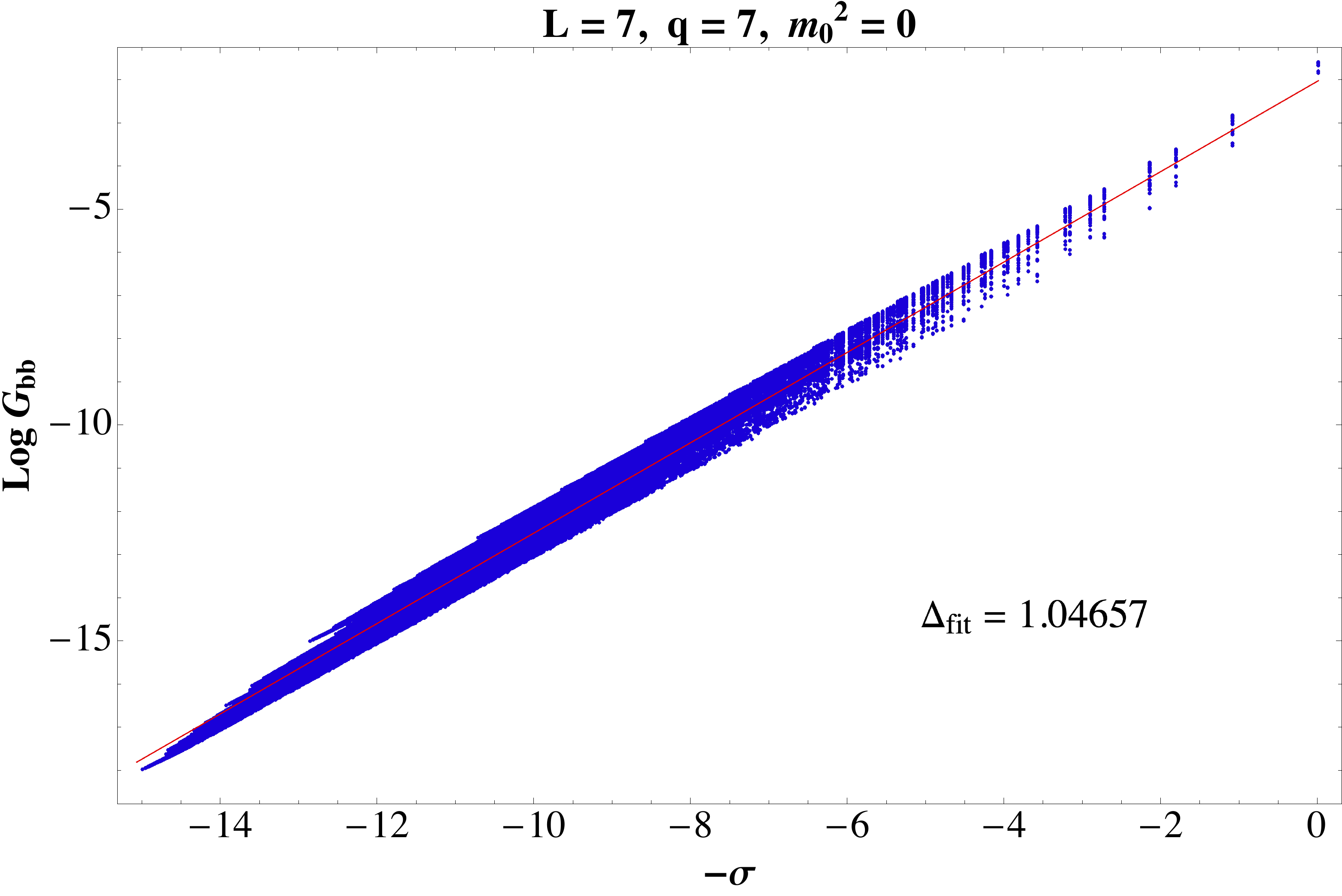}
\includegraphics[width=0.5\textwidth, angle=0,scale=0.99]{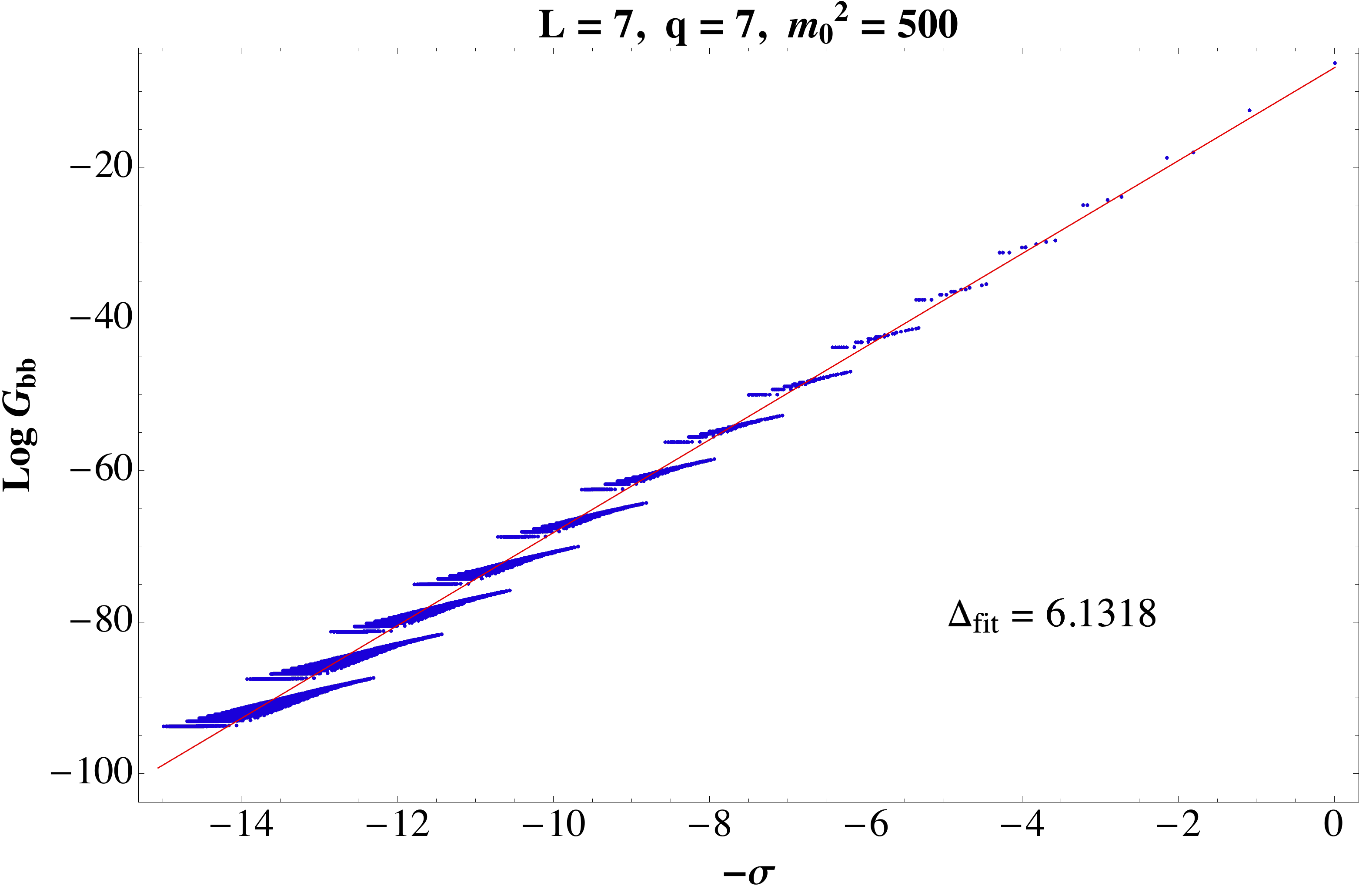}
\includegraphics[width=0.5\textwidth, angle=0,scale=0.99]{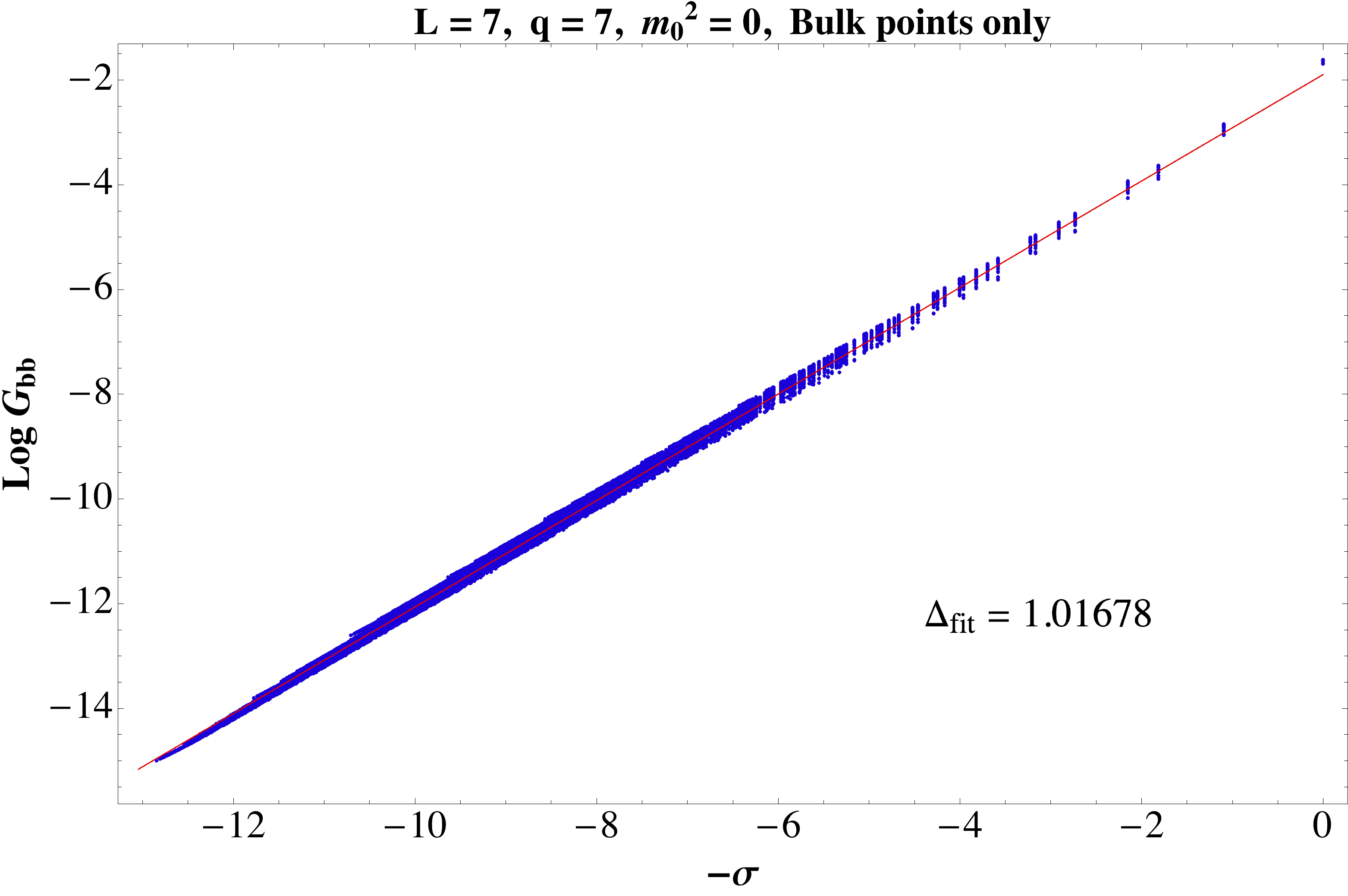}
\caption{A plot of (\ref{eq:log_eqn}) for different mass regimes. In the top plots the source point $X$ and sink point $X'$ cover all possible pairs on the lattice from which $\Delta_{\text{fit}}$ is derived from. In the bottom plot, only bulk points are used to avoid boundary effects. The linear fit is evident.}
\label{fig:deltalincheck}
\end{center}
\end{figure}

Recall the approximate formula for the continuum bulk-to-bulk propagator:
\be  \label{eq:log_eqn}
\log G_{bb}(\sigma) \approx  -\Delta \sigma. 
\ee
From the previous subsections, we expect that the discretized bulk-to-bulk propagator will behave similarly, but with a modified value for $\Delta$ compared to the usual continuum limit relation $m^2 \ell^2 = \Delta(\Delta-1)$.  Our strategy will therefore be to first check that $\log G$ is approximately linear in $\sigma$, and then from its slope extract $\Delta$. By doing this for many values of $m^2$, we obtain the function $\Delta(m^2$). Typical plots of $\log G$ as a function of $\sigma$ for fixed $m^2$ are given in Fig.\ \ref{fig:deltalincheck}. These show that for the lattice propagator, $\log G_{bb}(\sigma)$ is indeed linear in $\sigma$. 

Fig.\ \ref{fig:deltalincheck} also shows how only using bulk points in the linear fit to avoid boundary effects we can improve the derived $\Delta$. Additionally, in Sec.\ \ref{sec:ref} we show how using the exact fit from taking into account the true finite boundary condition produces a remarkably good estimate for $\Delta$ for a modestly size lattice. However, the approximate linear fit is sufficient for most of our purposes, as noted before.

The result of this comparison of the lattice propagator to our small and large mass approximate formulas ((\ref{eq:smallmassformula}) and (\ref{eq:largemassformula}), respectively)  is shown in Figs.\ \ref{fig:figdeltamass1} and \ref{fig:figdeltamass2}, showing an excellent agreement over a large range in mass.  

\FloatBarrier

\subsubsection{Approach to Cut-off Boundary}

\begin{figure}
\begin{center}
\includegraphics[width=0.75\textwidth, angle=0,scale=1.0]{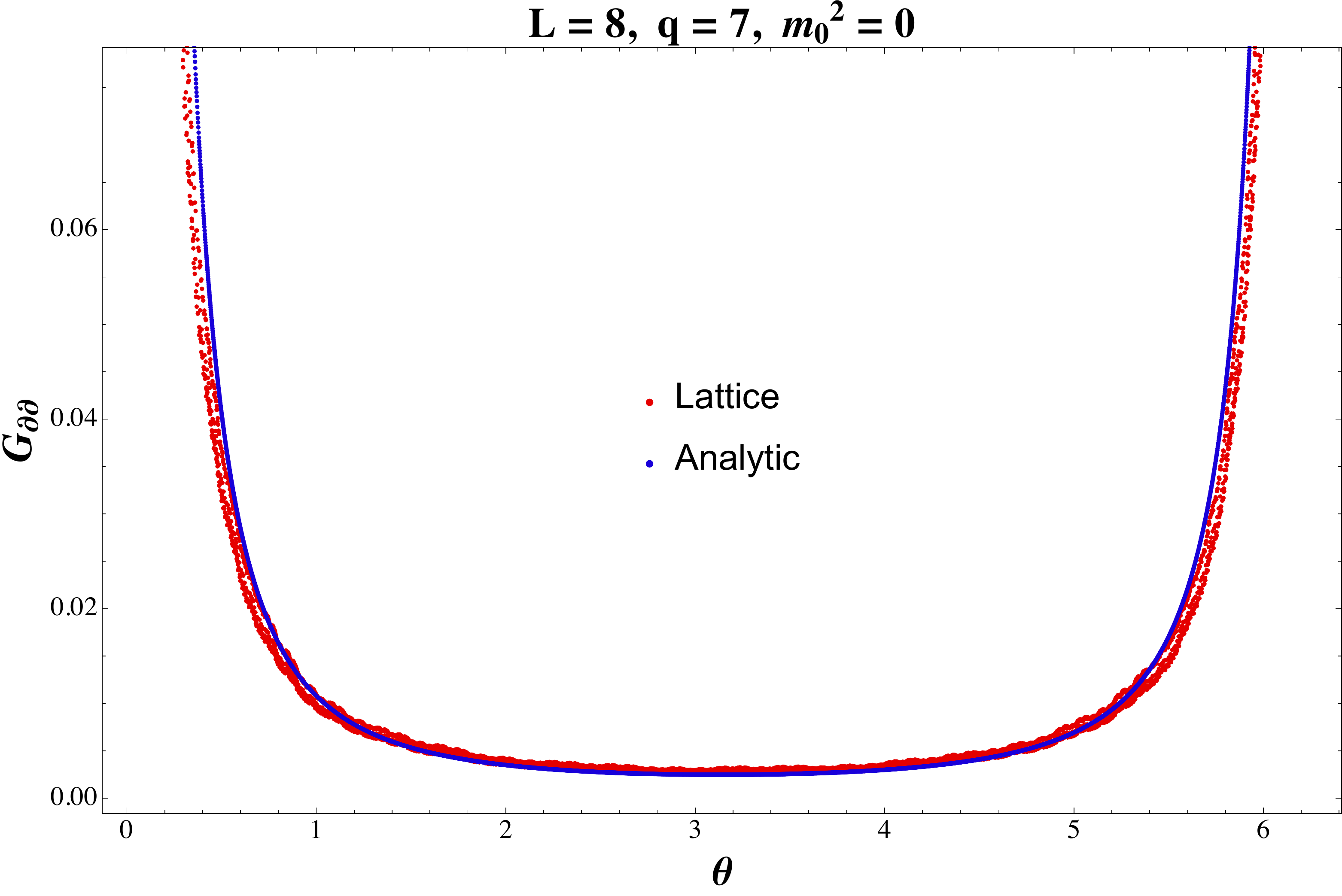}
\caption{Boundary-boundary two-point function from one point fixed on the boundary to all others, for lattice vs analytic continuum result.}
\label{fig:figboundbound}
\end{center}
\end{figure}

One subtlety in our tessellation scheme that we need to account for is that the lattice boundary points, although close, are not all at the same distance away from the origin on the Poincar\'e disk. This will affect comparing boundary dependent observables, such as the bulk-to-boundary propagator. We can see this by looking at the bulk-to-boundary propagator as the limit of the bulk-to-bulk propagator as the primed bulk coordinate approaches the boundary via $y' \rightarrow 0$:
\be
G_{b\partial}(y,x,x';\Delta) = \Big(\frac{y}{y^2 + |x - x'|^2} \Big)^{\Delta} = \lim_{y' \rightarrow 0}\frac{1}{y'^{\Delta}}G_{bb}(y,x, y', x'; \Delta) ,
\ee
where as $y' \rightarrow 0$ the divergent part goes as
\be
(y')^{-\Delta} \sim (1 - |z'|^2)^{-\Delta} .
\ee
So when looking at the bulk-to-boundary propagator from one point on the boundary, $X$, to any other point $X'$, we need
\be
G_{b\partial}(X,X';\Delta) \simeq (1 - |z'|^2)^{-\Delta}G_{bb,\,\text{lat}}(X,X'; \Delta) ,
\ee
where $G_{bb,\,\text{lat}}$ indicates the numeric lattice propagator data. In Fig. \ref{fig:figboundbound}, we compare the result from the lattice data to the exact CFT two-point function, which in our $\theta$ coordinate is 
\be
G_{\partial \partial}(\theta) \propto (1 - \cos\theta)^{-\Delta},
\ee
where the proportionality constant depends on the normalization of the operator, which we fix to match the lattice data.

\FloatBarrier

\subsection{Four-point Function}
\begin{figure}
\begin{center}
\includegraphics[width=0.75\textwidth, angle=0,scale=1.0]{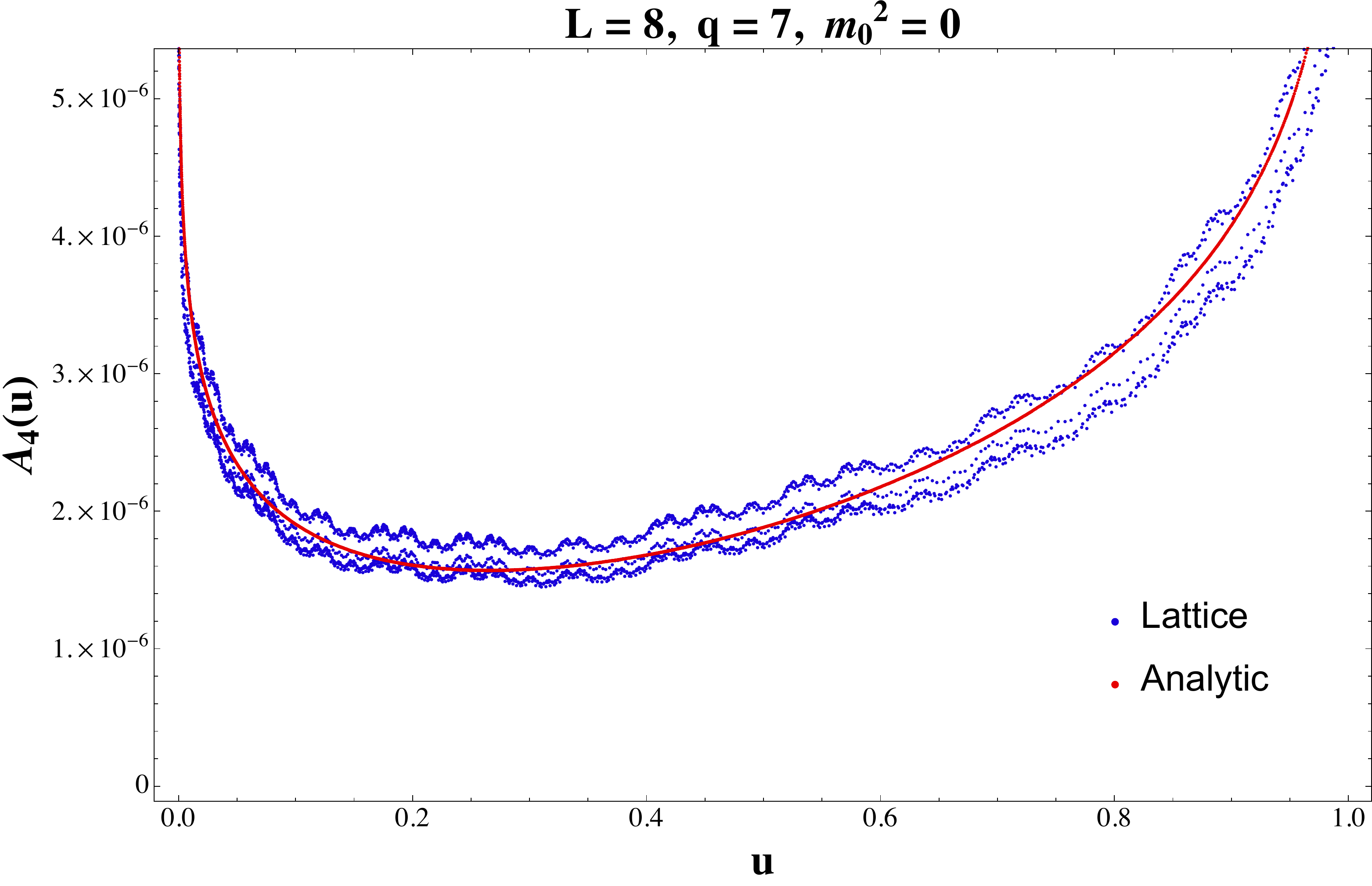}
 \caption{Four-point contact term from lattice vs analytic continuum result (\ref{eq:D1111}).}
\label{fig:fourpoint}
\end{center}
\end{figure}

Now that we have characterized the propagators we can calculate tree diagrams.
Here we focus on the four-point interaction term, which has a known analytic form. The AdS$_{2}$ four-point contact term is given by
\begin{equation}
A_{4}^{\text{Contact}}(x_i) =  D_{\Delta_1 \Delta_2 \Delta_3 \Delta_4} = \int d^{2}y \sqrt{g(y)} \, G_{b\partial}(y,x_1) G_{b\partial}(y,x_2) G_{b\partial}(y,x_3) G_{b\partial}(y,x_4)  .
\end{equation}
This becomes a sum on the lattice
\begin{equation}
A_{4}^{\text{Contact}}(x_i) \rightarrow \sum_{i} a^{\Delta}_{i} \, G_{b\partial}(x_i,x_1) G_{b\partial}(x_i,x_2) G_{b\partial}(x_i,x_3) G_{b\partial}(x_i,x_4)  ,
\end{equation}
where $a^{\Delta}_{i}$ is the area of the lattice triangle being summed over. Therefore we can easily calculate the four-point function by summing over four bulk-to-boundary propagator configurations. Moreover, for $\Delta_1 = \cdots = \Delta_4 = 1$ the $D$-function has a closed form given by
\be
\frac{2\,x^2_{13} x^2_{24}}{\pi^{d/2}} \, D_{1111}(x_i) = \frac{1}{z - \bar{z}} \bigg( 2 \Li_{2}(z) - 2 \Li_{2}(\bar{z}) + \log(z \bar{z}) \log \frac{1-z}{1- \bar{z}} \bigg)  ,
\label{eq:D1111}
\ee
which for $d=1$, $z = \bar{z}$, and to leading order becomes
\be
D_{1111}(x_i) \simeq \frac{2 x^2_{13} x^2_{24}}{\sqrt{\pi}} \bigg( -2 \, \frac{ \log (1 - z)}{z} + 2 \, \frac{\log z}{z-1} \bigg)  ,
\ee
allowing us to compare with the lattice value. The result of this comparison is shown in Fig.\ \ref{fig:fourpoint}, where we find a good agreement between the lattice and the analytic expression.

\FloatBarrier
\section{Finite Volume Corrections}
\label{sec:finitevolume}

In this section, we will discuss corrections to our discretization arising from the fact that any finite lattice fills out only a finite volume of AdS$_2$.  We will approach these corrections in two different ways.  The first is to compare our lattice results for the eigenvalues of the discretized Laplacian operator to the eigenvalues of a modified AdS$_2$ space where we move the boundary to a finite radius.  The second is to consider modifying our discretized action to take into account the effect of `integrating out' the rest of the space not included in our finite volume region.

\subsection{Eigenvalues of the Laplacian on Finite  Disk}

One measure of the accuracy of the lattice approximation is how closely the eigenvalues of the Laplacian are to the those of the continuum theory.  AdS$_2$ has infinite volume and the eigenvalues of its Laplacian form a continuous distribution.  However, for any finite number of layers, our tessellation fills out a finite volume, so it is more instructive to compare the eigenvalues of our discrete system to those of AdS$_2$ with a finite cutoff at $r=1-\epsilon$ away from the original boundary at $r=1$.  

To find the spectrum of the Laplacian, we first solve the equation of motion locally.  In global coordinates, 
take $f(r,\theta) = e^{i l \theta} f(r)$ with integer $l$.  The equation to solve is
\be
\nabla^2 f(r,\theta) = - \lambda f(r,\theta) \rightarrow \frac{\left(r^2-1\right)^2 \left(r \left(r f''(r)+f'(r)\right)-l^2 f(r)\right)}{4 r^2}=-\lambda  f(r).
\ee
Next, we impose the boundary condition that the solutions are regular at $r=0$. 
The regular solution is
\be
f(r) \propto r^l \left(1-r^2\right)^{\frac{1}{2} + i s} \, _2F_1\left(\frac{1}{2}+i s,l+\frac{1}{2} + i s;l+1;r^2\right),
\label{eq:eigenfunc}
   \ee
   where $s \equiv \sqrt{\lambda-\frac{1}{4}}$; although it is not obvious by inspection, the above function is invariant under $s\rightarrow -s$.
   The final step is to impose the boundary condition $f(1-\epsilon)=0$. This condition can be imposed numerically, and for any $\epsilon>0$ it is satisfied by the above solutions for a discrete set of values of $\lambda$.  At $r=1-e^{-y}$ in the limit of large $y$ ($r$ close to $1$) and small $s$, the eigenfunctions (\ref{eq:eigenfunc}) are approximately proportional to $\sin(s y)$:
   \be
   f(1-e^{-y}) \sim  \frac{e^{-y/2}}{\pi^{1/2} \Gamma(l+\frac{1}{2})}\left[ \frac{\sin s y}{ s} +\left(\log (2)-\gamma_E - \psi(l+\frac{1}{2})\right) \cos (s y) + {\cal O}(s)\right],
   \ee
   where $\psi(z)$ is the polygamma function and $\gamma_E $ is the Euler-Mascheroni constant. Therefore, with a boundary at very large $y= \log \frac{1}{\epsilon}$ the eigenvalues are approximately given by the discrete spectrum 
   \be
   s = \frac{n \pi }{y}\left( 1 + \frac{\gamma_E + \psi(l+\frac{1}{2})-\log (2)}{y} + {\cal O}(y^{-2}) \right), \qquad  n = 0,1,2, \dots.
   \ee
Since it is easy to numerically compute the exact spectrum of $s$ for any boundary value $\epsilon$ using the exact eigenfunction (\ref{eq:eigenfunc}), this is what we will use to compare to the lattice spectrum.

For the massless $L=7,\, q=7$ lattice the mean cutoff can be estimated as $\epsilon_{eff,7} \simeq 0.00082916$ 
by looking at the 
average value of $1-r$
of the \textit{next} $L=8$ layer where the Dirichlet BC is enforced. However, by looking at the spectrum we can do even better in defining a cutoff. 
In comparing the lattice and continuum eigenspectrums there are two parameters we can fix, the offset and the slope. The offset is fixed by the normalization between the lowest eigenvalues, which is $2.013$ in this case. We expect this to be close to the value given from the discrete Laplacian expansion at lowest order, $(7/4)(a^2/l^2) \simeq 2.081$. The slope is fixed by matching the second lowest eigenvalues. The result can be thought of as an improved definition for the cutoff, which in this case is $\epsilon_{eff,7} \simeq 0.00080897$.

\begin{figure}
\begin{center}
\includegraphics[width=0.75\textwidth, angle=0,scale=1.0]{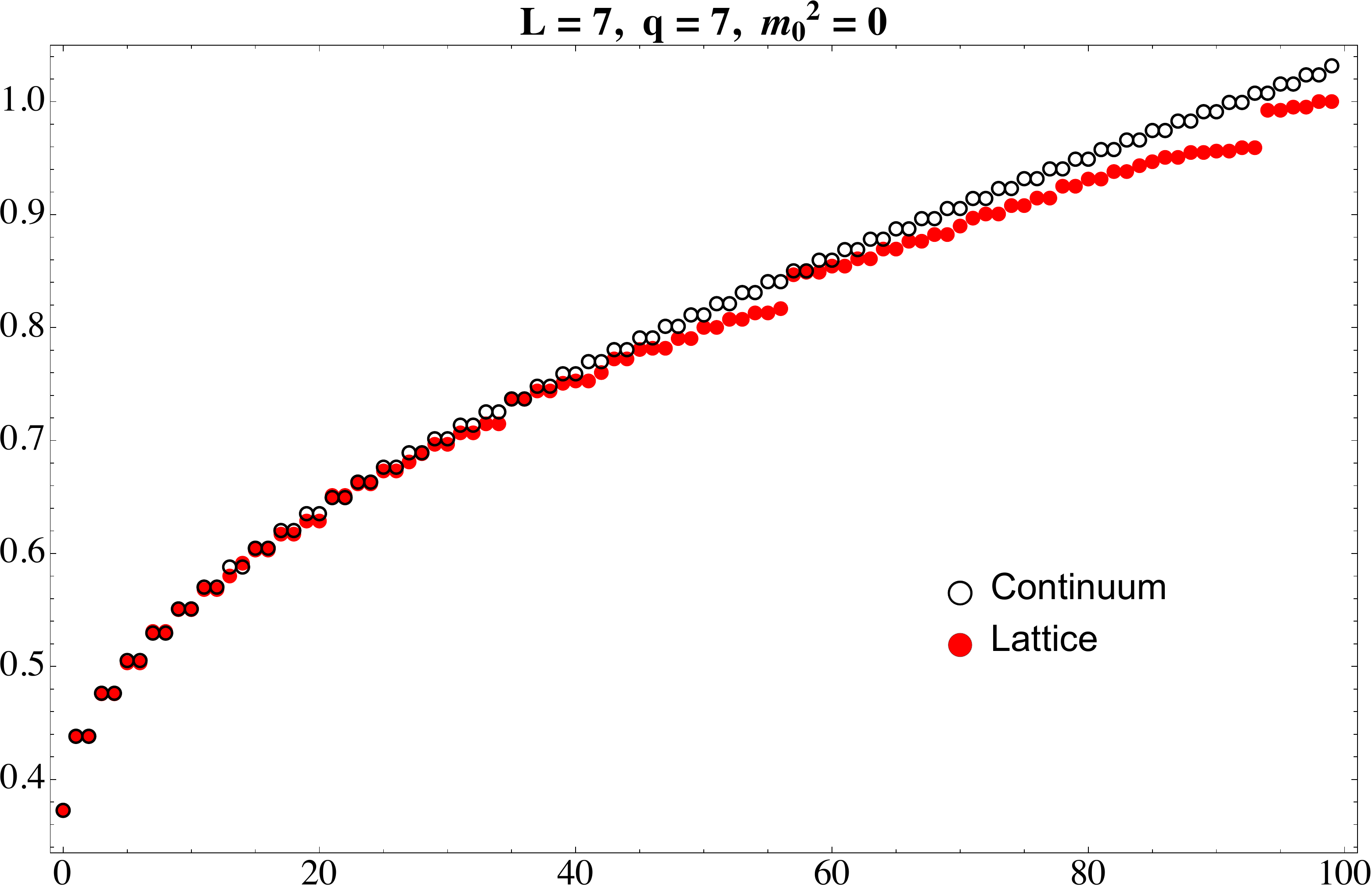}
\caption{The low-lying eigenspectrum of the continuum theory with a cutoff versus the discretized lattice realization. The normalization is fixed by matching the lowest eigenvalues, which is determined to lowest order by $c^2_{q}$. The slope is set by matching the second lowest eigenvalues. This can also be considered the definition of the cutoff.}  \label{fig:eigenspectrum}
\end{center}
\end{figure}

Fig.\ \ref{fig:eigenspectrum} shows a comparison of the lattice spectrum versus the spectrum from the continuum theory with the improved $\epsilon$ cutoff. The first 100 out of 4264 eigenvalues plotted show the low-lying spectrum; there is a remarkable agreement between the lowest eigenvalues of the continuum theory and the unrefined lattice.

\subsection{Integrating out the Boundary}

We can think of our lattice construction at finite $L$ as a low-energy effective theory realized from starting with the $L \rightarrow \infty$ limit and integrating out layers. Although the long-distance limit is associated with the IR, the modes of the dual CFT that live near the boundary are high energy ones, thus the Wilsonian intuition of ``integrating out" is still valid. This is an example of the UV/IR connection. By integrating out layers, we can gain another quantitative handle on how observable quantities depend on the number of layers of the lattice, especially in the large $L$ limit.

For a given $L$ the lattice Hamiltonian has the structure
\be 
H= 
\begin{pmatrix}
H_{b}  & \vline  & v & \\
\hline
 & \vline & &  \\
v^{T} &  \vline & H_{\partial} & \\
 & \vline & &
\end{pmatrix},
\ee
\\
where $H_{b}$ contain the bulk points, $H_{\partial}$ the boundary points, and $v$ and its transpose link boundary points to bulk points. This demarcation of the Hamiltonian naturally lends itself to a factorization of its determinant,
\be \label{eq:hamdet}
\det H = \det (H_{\partial})  \det (H/H_{\partial}),
\ee
where $H/H_{\partial} =  (H_{b} - v H^{-1}_{\partial} v^{T})$ is the Schur complement of $H$ relative to $H_{\partial}$. The importance of this term is that it encodes the precise correction to the bulk from the boundary layer.

\begin{figure}
\begin{center}
\includegraphics[width=0.75\textwidth, angle=0,scale=1.0]{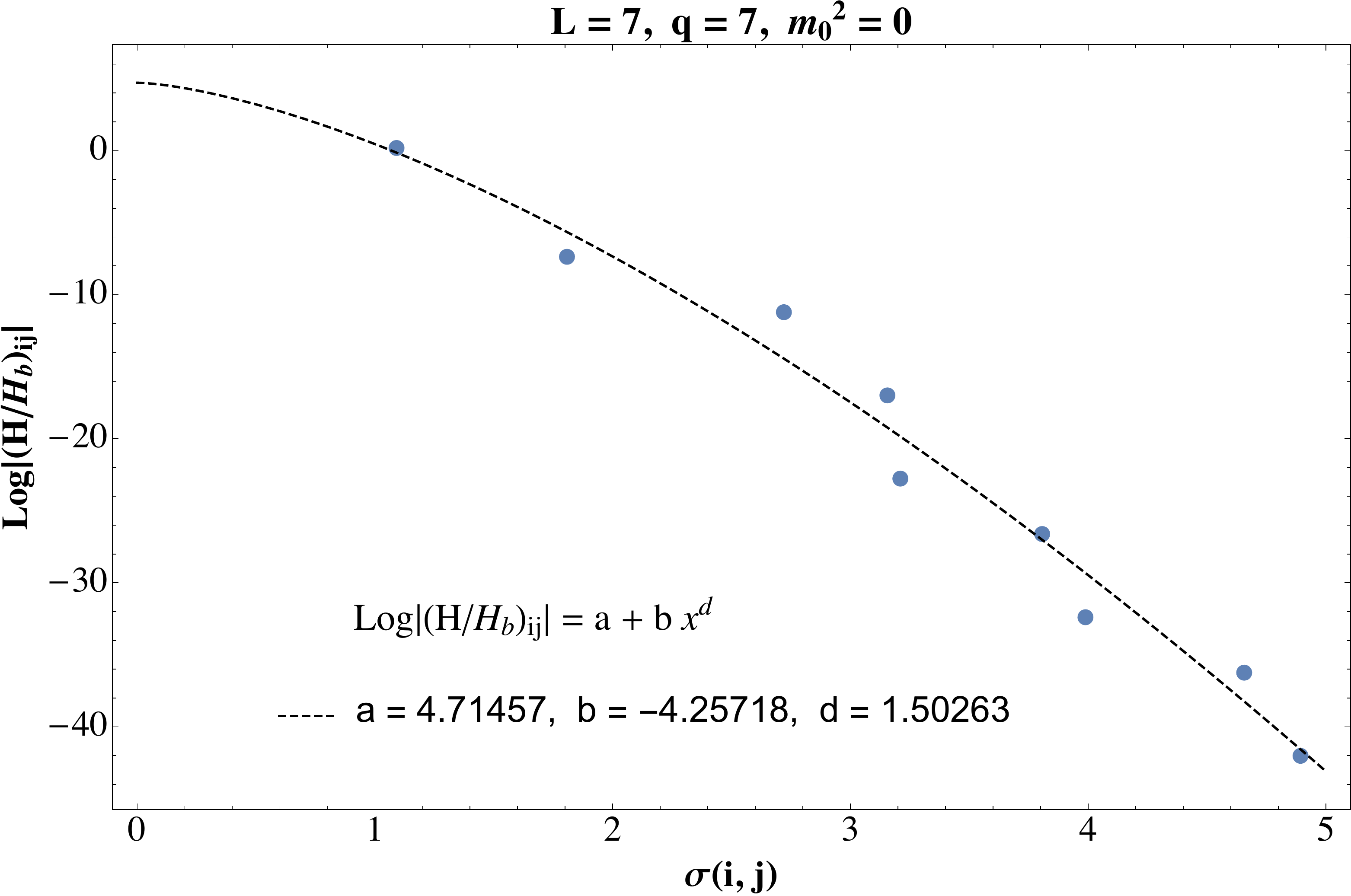}
\caption{The value of the matrix element of a chosen boundary point $i$ in the Schur complement $H/H_{\partial}$ and its boundary neighbors $j$. The exponential decay signals the vanishingly small coupling between non-local boundary points as a result of adding a new layer.}  \label{fig:geoboundarylocality}
\end{center}
\end{figure}

Whereas $H_b$ is manifestly local, the corrective piece $v H^{-1}_{\partial}v^{T}$ in the Schur complement couples non-local points. That is, given a point on the boundary which corresponds to some matrix element $(H/H_{\partial})_{ii}$ its non-neighboring elements $(H/H_{\partial})_{i\,i+2}$, $(H/H_{\partial})_{i\,i+3}$, ... are non-zero. Now, before integrating out the $(L+1)$-th layer, all points strictly have nearest-neighbor links, so one boundary point will only be linked to its two boundary neighbors and the nearest-neighbor bulk points. Moreover, the two-point function for a particle propagating along the boundary is given by $G_{\partial \partial} \propto (1-\cos \theta)^{-\Delta}$. Consequently,  we expect the non-locality arising from integrating out the extra layer to be suppressed as a function of the geodesic distance. Indeed, as Fig.\ \ref{fig:geoboundarylocality} shows, by picking a boundary point and looking at its coupling to other boundary points, this non-local effect decays exponentially in geodesic distance. 

In summary we have shown that for any lattice size $L$ we understand precisely how the $L+1$ layer corrects the Hamiltonian and that this correction preserves locality, so we can view our lattice as a local, low-energy effective theory obtained by integrating out layers from the $L \rightarrow \infty$ continuum.

\section{Finite Lattice Spacing Refinement} \label{sec:ref}
 
So far we have looked at the pure equilateral triangle group
tessellation as an approximation of the hyperbolic plane. Relative to the
continuum this introduces a finite lattice spacing error, with  $(2,3,7)$
triangles having a  minimum spacing of $a \simeq 1.091$ relative to the fixed curvature length $\ell$.
The finite element approach to  partial differential equations  uses a sequence of nearly regular
smaller simplices (triangles in 2d) when properly constructed allowing one to converge
to exact  continuum solutions, or in the context of quantum field theory to
the classical tree approximation. Loops introduce UV divergences 
\cite{PhysRevD.98.014502} but are neglected in our current  
presentation.

The linear interpolated finite elements method in 2d flat space for 
the Laplace Beltrami operator is equivalent to the method of
discrete exterior calculus (DEC)~\cite{2005math8341D}.
The formal expression
for the DEC Laplace-Beltrami operator acting on scalar fields in d-dimensions is
$(\bm{\delta}  +   {\bf d}  )^2\phi = ({\bf d} \bm{\delta}  +   \bm{\delta}{\bf d}  )\phi = *  {\bf d} * {\bf d} \phi$, with  
\be
* {\bf d} * {\bf d} \phi(i) = \frac{|\sigma_0(i)|}{|\sigma^*_0(i)|} \sum_{j \in\<i,j\>} 
                      \frac{|\sigma^*_1(ij)|}{|\sigma_1(ij)|} (\phi_i
                      -\phi_j) = \frac{1}{\sqrt{g_i}} \sum_{j \in \<i,j\>} \frac{ S_{ij}}{l_{ij}} (\phi_i
	-\phi_j)  
\label{eq:LB}
\ee
in terms of the discrete exterior derivative ${\bf d}$ and its 
dual $\bm{\delta} = *d*$. The notation
$\sigma_n$ identifies a $n-$dimensional simplex~\footnote{More precisely,
$\sigma_n(i_0,i_1,\cdots i_n)$ is the  volume form for the n-simplex
with vertices $i_0,i_1,\cdots i_n$ with value $\pm|\sigma_n|$  for even/odd 
permutations of the indices. See Ref.~~\cite{2005math8341D} for details. }
\verb!points, lines, triangles,! etc. for $n = 0,1,\cdots, d$ and
$|\sigma^*_n|$  the $d-n$ dual polyhedron. $| \cdots | $ is the volume
for these polyhedra.  By convention $|\sigma_0(i)| =1$. This coincides with the natural form of the divergence as
a flux, where $S_{ij}$ is the volume of the dual normal.  To be concrete, in Fig. \ref{fig:latanddual} (where $l_{ij} = |\sigma_1(ij)|$ is also indicated), $S_{ij}$ is the length of the link $l^*_{ij} = |\sigma^*_1(ij)|$.

On our hyperbolic manifold it is natural and convenient to replace 
$l_{ij}$ and $l^*_{ij}$ by the appropriate geodesic lengths as we have already been
using this intrinsic geometry for the triangle group. To do this we need
Heron's hyperbolic triangle area rule and the circumradius~\cite{horvath:2014},
\bea
A(a,b,c) &=& 
 4 \arctan \bigg[ 
   \sqrt{\tanh\frac{s}{2} \tanh\frac{s - a}{2} \tanh\frac{s - b}{2} \tanh\frac{s-c}{2}} \bigg] \nn
R(a,b,c) &=& 
\arctanh\bigg[ \Big( \tanh\frac{a}{2} \tanh\frac{b}{2} \tanh\frac{c}{2} \Big)/\sin\frac{A(a, b, c)}{2} \bigg],
\eea
for a general hyperbolic triangle $\triangle(a,b,c)$ in terms of the side lengths $a,b,c$, with $s = (a + b+
c)/2$.  Finally, using the
Pythagorean theorem for a hyperbolic right triangle, 
$\cosh (c)  =  \cosh (a) \cosh (b)$, we can solve for the length of the dual\footnote{The ``dual from a side $a$ into the triangle $a,b,c$'' means the segment from the center of $a$ to the circumcenter of triangle $a,b,c$.}  from a side $a$ into the triangle
$a,b,c$ -- denoted $l^{*}(a,b,c)$ -- as
\be \label{eq:dualformula}
l^*(a,b,c) = \arccosh \big[ \cosh R(a,b,c) / \cosh(a/2) \big].
\ee
The above formulae allows us to calculate the the weights $l^{*}_{ij}/l_{ij}$ on an arbitrary lattice. 

For the first iteration of this hyperbolic refinement we
split our fundamental equilateral triangle into four sub-triangles by
bisecting each edge to half the lattice spacing: $a \rightarrow a/2$. 
The price we pay is a breaking of the $\mathbb{Z}_7$ symmetry at this scale
so the sub-triangles are neither equilateral nor congruent to 
each other. This means the FEM weights are no longer uniform.

\begin{figure}
\begin{center}
\includegraphics[width=0.4\textwidth, angle=0,scale=0.81]{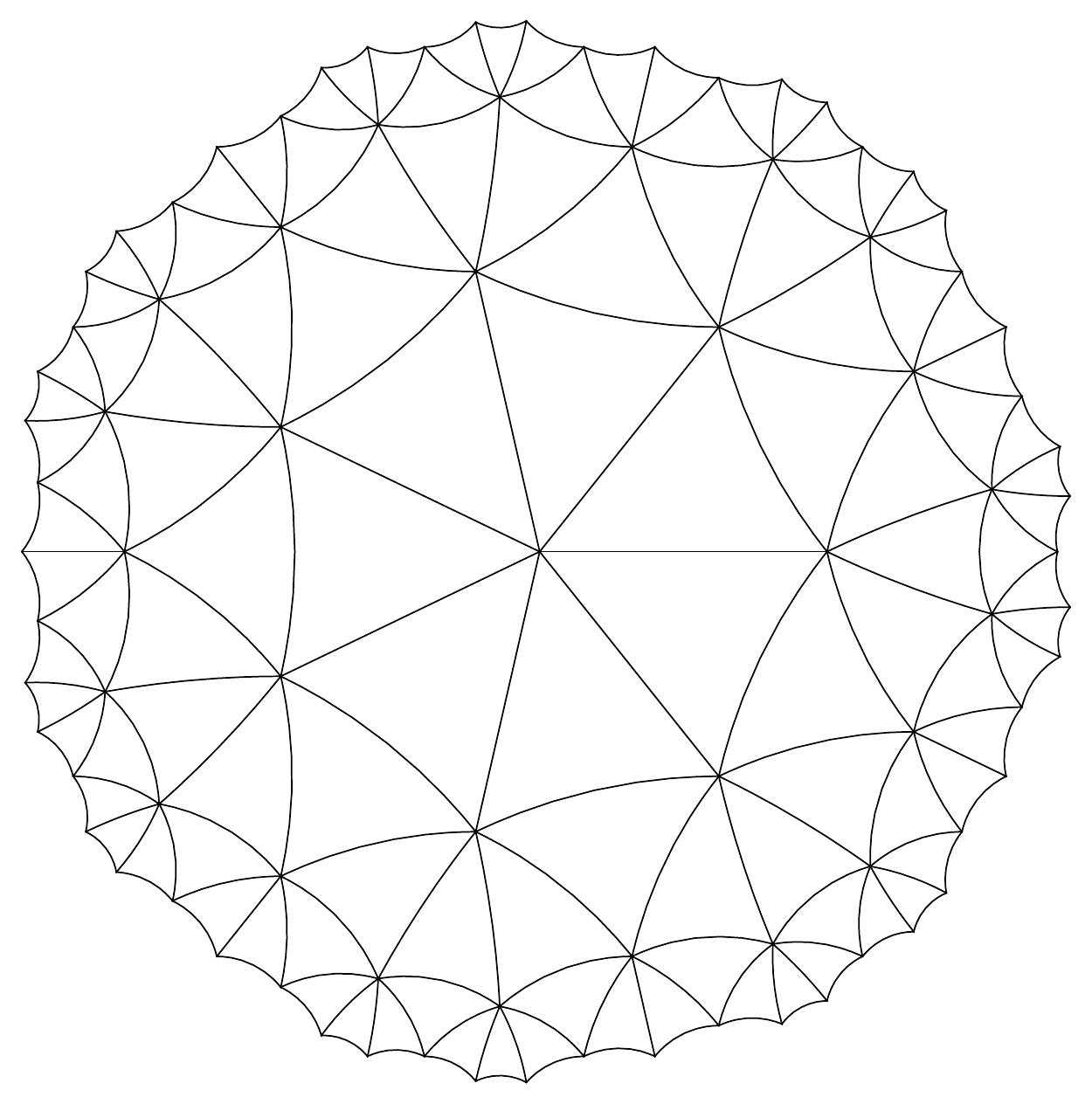}
\includegraphics[width=0.4\textwidth, angle=0,scale=0.81]{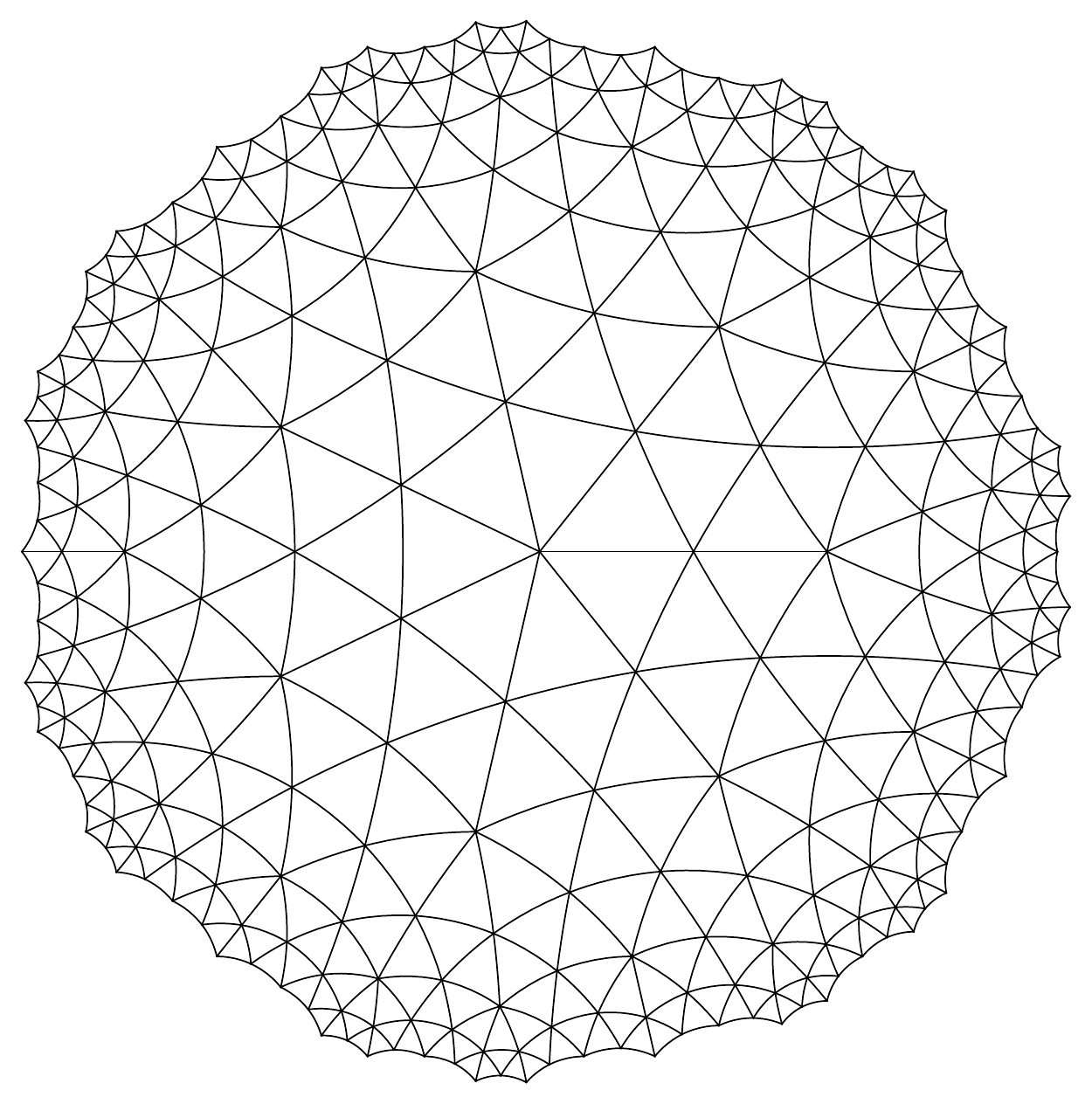}
\includegraphics[width=0.4\textwidth, angle=0,scale=0.81]{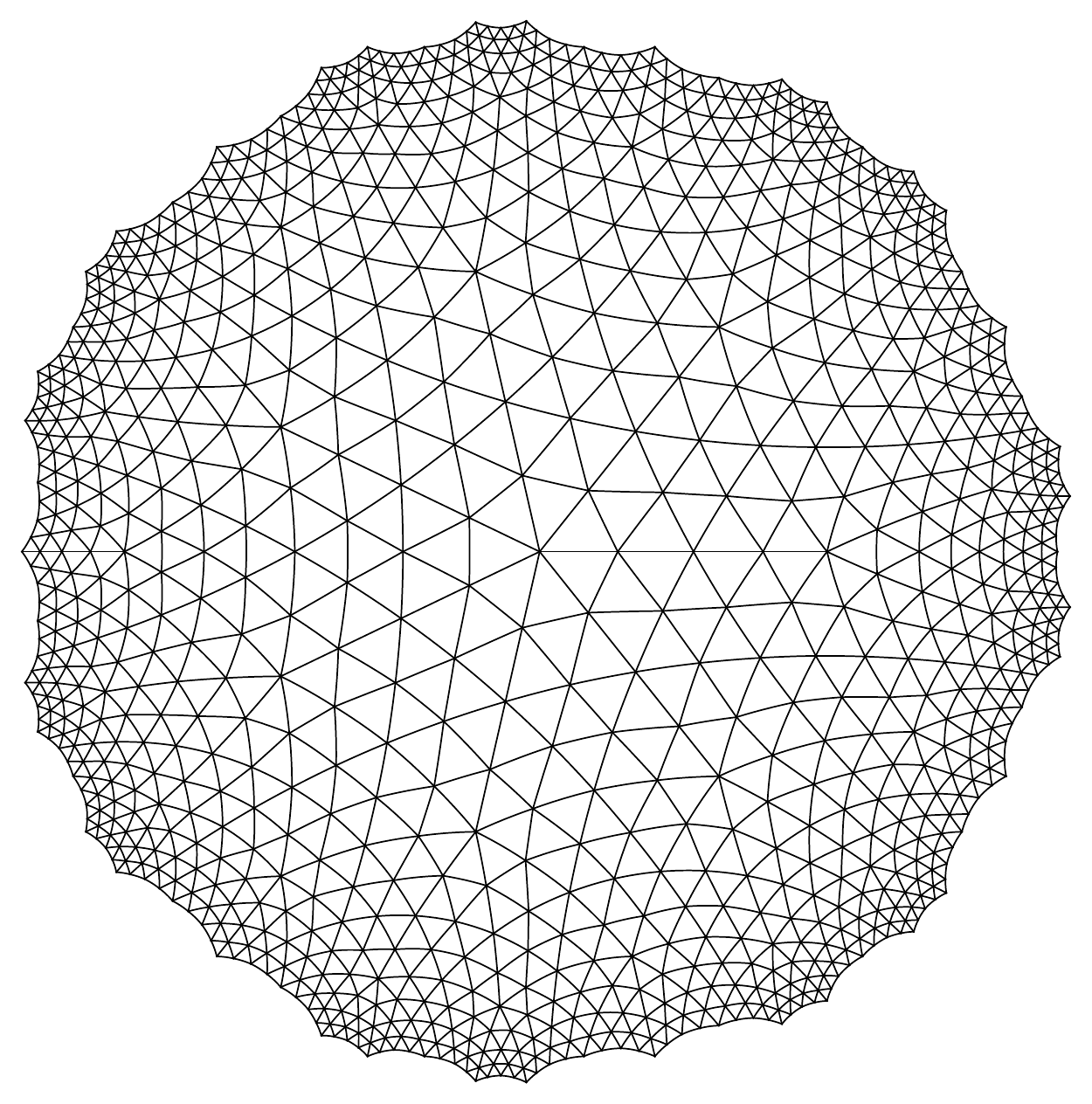}
\caption{Refinement of the $L = 3$, $q = 7$ lattice. The lattice is unrefined on the left, has one refinement in the middle, and two refinements on the right. }  \label{fig:latticerefplot}
\end{center}
\end{figure}

Explicitly, the equilateral triangle at one refinement is split into three triangles of side lengths
$(a/2,  a/2,  c)$ and one with sides $(c, c, c)$, where for $q=7$
\be
a = 2 \arccosh \Big[\cos\frac{\pi}{3} \Big/ \sin\frac{\pi}{7}\Big] \approx 1.091, \quad 
c = \arccosh \Big[\cosh^2 \frac{a}{2} - \sinh^2 \frac{a}{2} \cos \frac{2\pi}{7} \Big] \approx 0.492.
\ee
Using (\ref{eq:dualformula}) we find the kinetic weight $l^{*}_{ij}/l_{ij}$ between two points $i$ and $j$, which, is the factor $K_{ij}$ in (\ref{eq:lattice_action}).  For a single refinement there are only two different weights, one for those links that are already present in the original lattice (left image in Fig. \ref{fig:latticerefplot}), and another for those that are new in the refined lattice (middle image in Fig. \ref{fig:latticerefplot}).  These two weights are 
\be
K_{ij, \, \rm orig} = \frac{2l^*(\frac{a}{2}, \frac{a}{2},c)}{a/2} = 0.490, \qquad K_{ij, \, \rm ref} = \frac{l^*(c,c,c)+l^*(c,\frac{a}{2},\frac{a}{2})}{c} = 0.643.
\ee
Note that for $K_{ij, \, \rm ref}$, the length $l_{ij}^*$ is the sum of two different lengths, one for the normal in each direction. 

\begin{figure}
\begin{center}
\includegraphics[width=0.5\textwidth, angle=0,scale=0.99]{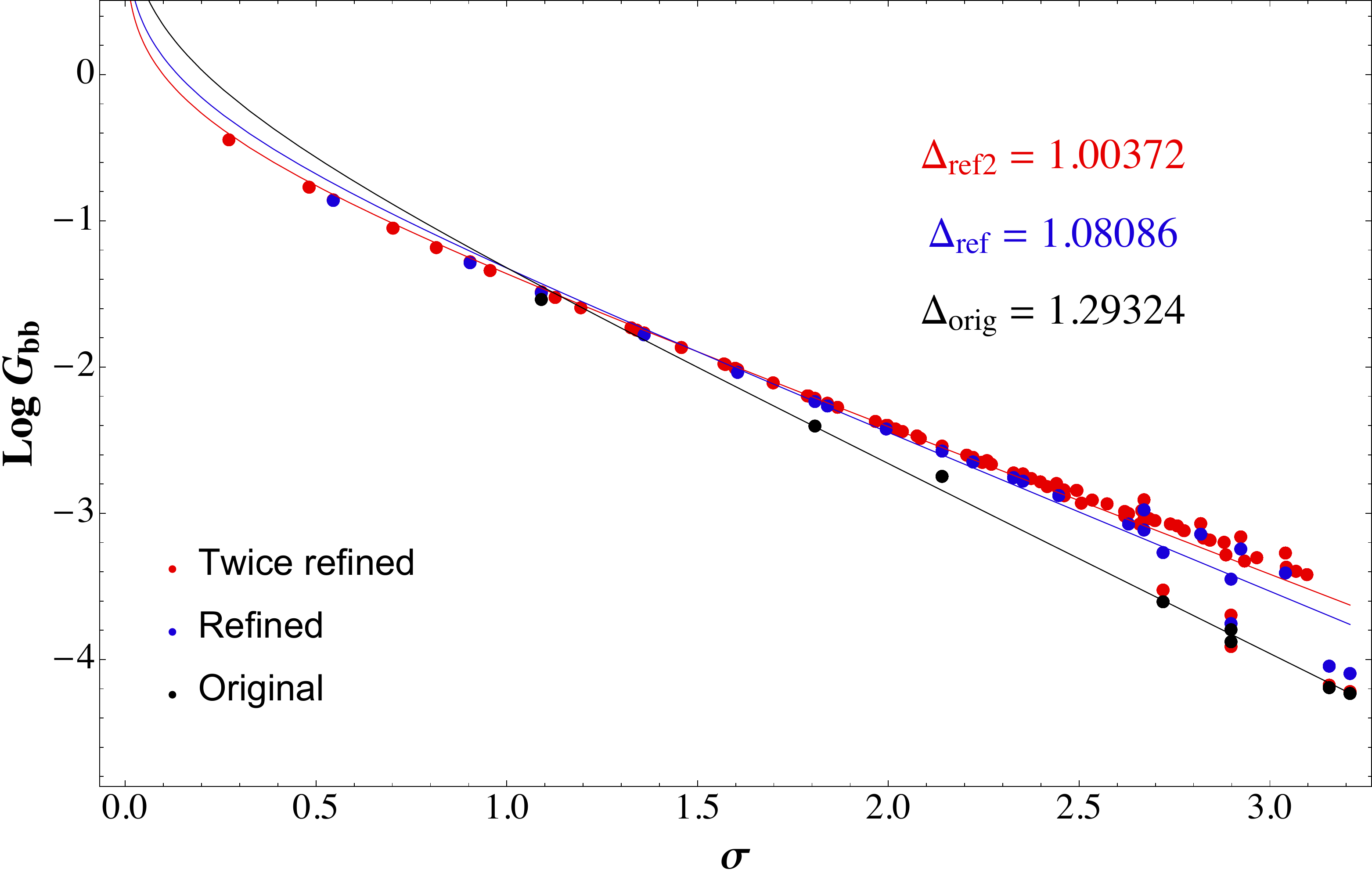}
\includegraphics[width=0.5\textwidth, angle=0,scale=0.99]{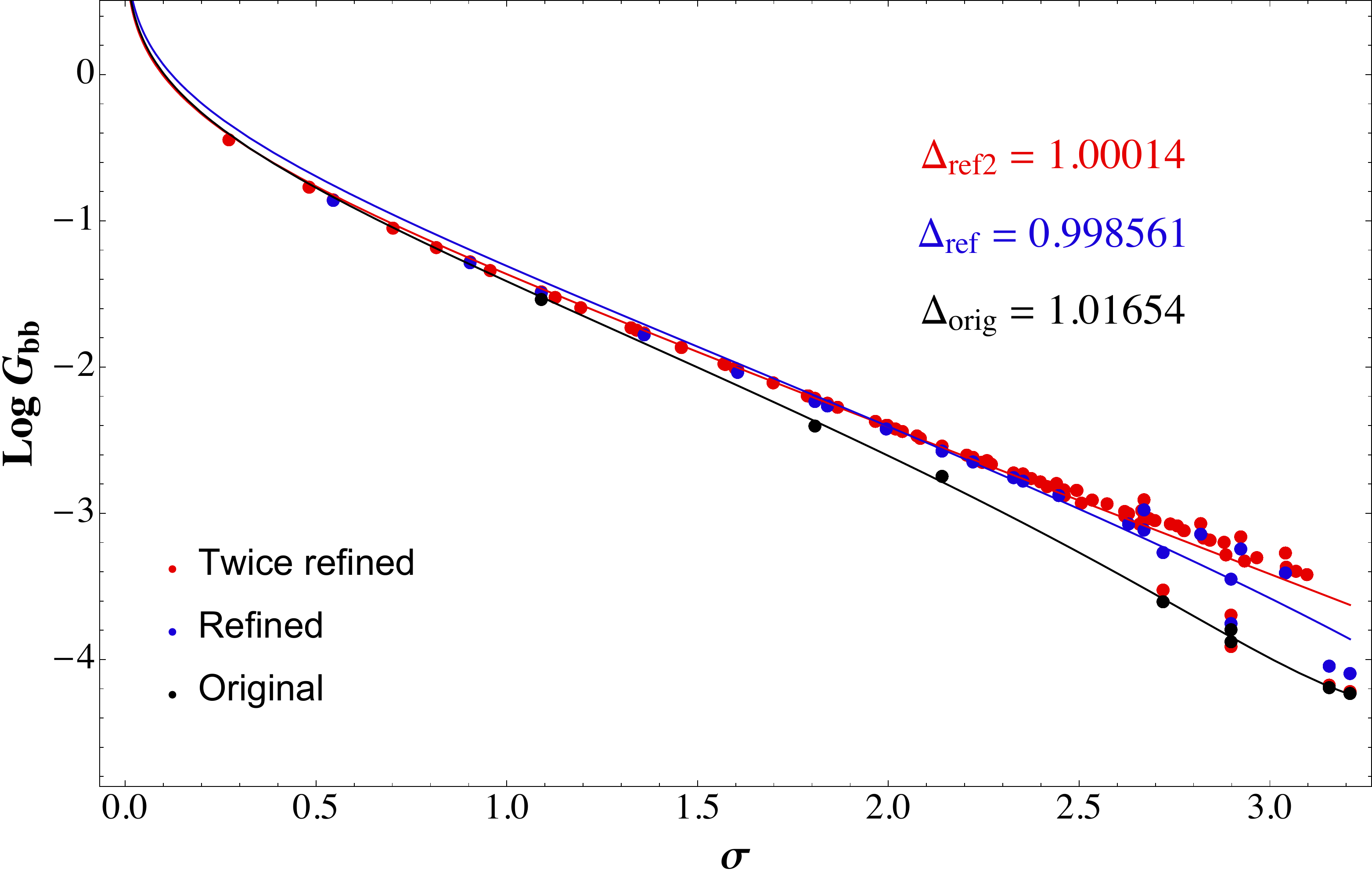}
\caption{The bulk-to-bulk propagator from a central source to all other points for the massless $L=3$, $q=7$ unrefined, once refined, and twice refined lattice, respectively. The fit for $\Delta$ is derived from the infinite BC corresponding to (\ref{eq:exactbulkbulk}) for the left figure and solving the exact equation for a finite BC on the right. }
\label{fig:bbref}
\end{center}
\end{figure}

The rightmost image in Fig.\ \ref{fig:latticerefplot} shows an example of a lattice refinement for two refinements. In principle this process can be applied recursively an arbitrary number of times. For the case of a lattice with a finite number of layers, in the limit of an infinite number of refinements we recover the continuum theory with a cut-off. More concretely, observables such as the bulk-to-bulk propagator that are sensitive to short distances in the bulk 
become closer to their continuum values
as we add more refinements. Fig.\ \ref{fig:bbref} shows this is indeed the case. We emphasize that Fig.\ \ref{fig:bbref} also shows that by using the correct boundary conditions, the unrefined triangle group lattice -- even with a modest number of layers -- produces remarkably accurate results.

\FloatBarrier

\section{Future Directions}
\label{sec:future}

We have presented a triangulation of  AdS$_{2}$ in order to establish
a framework for non-perturbative lattice field theory methods on the hyperbolic
disk $\mathbb H^2$.  The $D(2,3,q)$ triangle
group generates a lattice of equilateral triangles that preserve an exact
discrete subgroup of the isometries.  The case $q = 7$
gives the minimum area triangulation with a lattice spacing $a$ that is ${\cal O}(1)$ in units of the curvature $K = 1/\ell^2$.
Nonetheless,  when  introducing $\phi^4$ theory we find 
remarkably good comparisons for propagators, the tree-level four-point
amplitude, and the Laplacian eigenspectrum on a Poincar\'e disk of finite geodesic
radius. We also introduce methods to reduce the lattice spacing by
application of finite elements refinement using discrete
exterior calculus. We propose that this initial lattice is sufficient for a first
semi-quantitative, non-perturbative investigation of the AdS$_2$/CFT 
correspondence in the limit that gravity is decoupled.

We have restricted our attention to a scalar field theory in AdS$_2$ for simplicity, but it would be interesting to generalize to a wider class of field theories. 
 The inclusion of bulk Abelian and non-Abelian gauge theories
and Dirac Fermions should be straightforward following the approach of
Ref.~\cite{Brower:2016vsl} for de Sitter.  K\"{a}hler Dirac fermions are
naturally adapted to simplicial lattices using  DEC as described in
Ref.~\cite{Catterall:2018dns}.  Generalizing to higher dimensions should also be possible, and it is an interesting question what constitutes the best choice for lattices on higher dimensional AdS.

Ultimately, the point is to eventually do nonperturbative Monte Carlo lattice simulations of bulk field theories, and compute correlators in the dual boundary CFTs.  A natural first effort is to study $\lambda \phi^4$ theory AdS$_2$ at strong coupling.  In that case, one could compare results for the spectrum of the dual CFT$_1$ to constraints from the conformal bootstrap \cite{Paulos:2016fap}.
Although we do not expect to be able to include gravity nonperturbatively with a lattice computation, it may be possible to include it in the perturbative gravity limit.  Including such corrections would improve the dual CFT to include a stress tensor, albeit one with a large central charge.
The Regge calculus approach, which can sum over weak metric fluctuations, may be naturally suited to this task.

We also note that hyperbolic graphs are playing an increasingly prominent
role in other applications such as tensor
networks~\cite{Evenbly:2017hyg, Jahn:2017tls, Boyle:2018uiv} and
quantum error correction codes~\cite{Swingle:2009bg,Almheiri:2014lwa}.
These networks are employed
in the computation of entanglement
  entropies~\cite{PhysRevB.99.155126} for boundary field theories.  For this reason, our sketch of the triangle group
algebra  may have relevance beyond the present context.

\section*{Acknowledgements}
We would like to thank Simon Catterall for enlightening and helpful discussions. ALF and RCB were supported in part by the US Department of Energy Office of Science under Award Number DE-SC0015845. ALF was supported in part by a Sloan Foundation fellowship. This work was supported in part by the U.S. Department
of Energy (DOE) under Award No. DE-SC0019139.


\appendix
\section{Hyperbolic Plane}
In this appendix we reproduce various results of the hyperbolic plane. 

\subsection{Metric}
\label{app:metric}
There are various forms of the metric:
\begin{align} \label{eq:hyp_metric}
ds^2 &= \ell^2 (d\rho^2 + \sinh^2 \rho d\theta^2) ~~~~~~~~~~(\text{Geodesic coordinates}) \\
&= \frac{4 \ell^2}{(1-r^2)^2} (dr^2 + r^2 d\theta^2) ~~~~~~~(\text{Global coordinates: }r=\tanh \frac{\rho}{2},~\rho=\log \frac{1+r}{1-r}) 
\label{eq:glob_metric}\\
&= 4 \ell^2 \frac{dz dz^{*}}{(1-|z|^2)^2} ~~~~~~~~~~~~~~~~~~~ (\text{Poincar\'{e} disk: } z=r e^{i\theta}) \\
&= \ell^2 \frac{dw dw^{*}}{(\Im w)^2} ~~~~~~~~~~~~~~~~~~~~~~~~  (\text{Upper-half plane: } z=\frac{w-i}{w+i}) \\
&= \ell^2 \frac{dx^2 + dy^2}{y^2} ~~~~~~~~~~~~~~~~~~~~~  (\text{Upper-half plane: } w=x+iy) 
\label{eq:UHPMetricXY} \\
& = \ell^2 \left(-dX_0^2 + dX_1^2 + dX_2^2 \right) ~~~~ (\text{Embedding Space: } X_0=\cosh \rho, X_1 \pm i X_2 =  e^{ \pm i\theta}\sinh \rho ) 
\end{align}
The chordal distance 
\be
d_{\rm ch}(X,X')\equiv -\frac{1}{2}(X-X')^2 \equiv \ell^2 \left(1- \frac{1}{\xi(X,X')} \right), \qquad \xi(X,X') \equiv -\frac{\ell^2}{X\cdot X'},
\ee
is related to the geodesic distance (in units of AdS radius $\ell$) $\sigma(X,X')$ by
\be
\cosh(\sigma) = \frac{1}{\xi} \quad \leftrightarrow \quad \sigma = \log \left( \frac{1+ \sqrt{1-\xi^2} }{\xi} \right),
\ee
as is easily verified by evaluating the chordal distance in geodesic coordinates between the point at $\rho=0$ and any other point.  The embedding space coordinates are related to the UHP coordinates by
\be
X_0 = \frac{1+ x^2 + y^2}{2y}, \quad X_1 = \frac{-1 + x^2 +y^2}{2y}, \quad X_2 = -\frac{x}{y},
\ee
from which one can easily obtain
\be
\xi(X,X') = \frac{2 y y'}{(x-x')^2 + y^2 + y'^2}.
\ee

Lastly, we also mentioned that, for the {Poincar\'{e} disk}, the geodesic distance  between two points $z$ and $z'$ can be expressed as
\be
\sigma (z,z') = 2 \tanh^{-1} \Big| \frac{z-z'}{1-z^* z'}\Big | .
\ee

\subsection{Hyperbolic Trigonometry} \label{app:hyptrig}

Now we turn to triangles, with interior angles $(\alpha, \beta, \gamma)$ with opposite geodesic side lengths of $(a,b,c)$ in units of curvature $\ell=1$. We will be primarily interested in the case of negative constant curvature, where $\alpha+ \beta +\gamma <\pi$. The triangle area   $A_\triangle$  is given by (\ref{eq:HypTriangleArea}), with ratios for its sides fixed.  
For equilateral triangles, $\alpha=\beta=\gamma$ and $a=b=c$,  one has
\be
\cosh a/2 = \frac{1}{2\sin \alpha/2}.
\ee
In this paper we are interested primarily in equilateral triangles with $\alpha=\beta=\gamma=\frac{2\pi}{q}$ and $q=7, 8, \cdots$, thus
\be
a=2\cosh^{-1}\frac{1}{2\sin \pi/q}.  \label{eq:equilateral}
\ee
For right triangles with $\gamma=\pi/2$,
\be
\cosh a =\frac{\cos \alpha}{\sin \beta}, \quad \cosh c = \frac{\cos \beta}{\sin \alpha}, \quad \cosh c = \cot \alpha \cot \beta .
\ee
The last relation corresponds to the  Pythagorean theorem. 
In particular, consider  $(\alpha,\beta,\gamma)=(\frac{\pi}{2},\frac{\pi}{3}, \frac{\pi}{q})$.  For the side opposite the angle $\beta=\pi/3$, 
\be
b=\cosh^{-1} \frac{\cos \pi/3}{\sin \pi/q}= \cosh^{-1} \frac{1}{2\sin \pi/q} = a/2 ,
\ee
where $a$ above is the geodesic length for the side of the $(q,q,q)$ equilateral triangle, given in (\ref{eq:equilateral}), and also in (\ref{eq:latticeLength}).

\section{Action of Triangle Group on the Lattice} \label{app:trigrouplattice}
 
The maximally symmetric lattices we use in this paper are irreducible representations of the proper  $D(2,3,q)$ triangle group, so that every point in the lattice is related to every other point by a symmetry of the lattice.  Here, we exhibit the action of this symmetry explicitly, in Poincar\'e disk complex coordinates $z = r e^{i \theta}$  mapped from the UHP coordinates $w = x +iy$ by $z = (w -i)/(w +i)$.
 The symmetry of the continuum spacetime is $PSL(2, \mathbb{R})$, of which the triangle group is a discrete subgroup.  

As discussed in Section \ref{sec:Triangle}, the triangle group $D(p,r,q)$ is generated by two elements $S$ and $T$ satisfying
\be
S^p = 1, \quad T^r = 1, \quad U^q = 1, \quad U= ST.
\ee
We are interested in the case $(p,r,q)=(2,3,q)$ so that six triangles with angles $(\frac{\pi}{2}, \frac{\pi}{3}, \frac{\pi}{q})$ form a cell of an equilateral triangle, as illustrated in Fig.\ \ref{fig:equitri}, and our lattice is the vertices of these equilateral triangles. We will orient our lattice so that one  
vertex is at the center $z=0$ in Poincar\'e disk coordinates, and a neighboring  vertex  is along the positive real axis as in  Fig.~\ref{fig:latanddual}. 
The action of $U$ on the lattice is just a rotation by $2\pi/q$:
\be
U: z \rightarrow e^{2 \pi i/q} z.
\ee
Using the map $z = \frac{w-i}{w+i}$ from the Poincar\'e disk to the UHP, it is easy to infer the action of $U$ on the lattice in UHP coordinates.  We can represent this action as an element of $PSL(2,\mathbb{R})$ as follows:
\be
U: w \rightarrow \frac{a w+b}{c w+d}, \qquad  \left( 
\begin{array}{cc} a & b \\ c & d \end{array} 
\right) = 
 \left( 
\begin{array}{cc} \cos \frac{\pi}{q} & - \sin \frac{\pi}{q} \\
 \sin \frac{\pi}{q} & \cos \frac{\pi}{q} \end{array} 
\right) .
\ee
The action of $S$ and $T$ are more difficult to infer, but easy to check once they are known.  For $S$, its action on the UHP is given by
\be
S \cong \left( \begin{array}{cc} 0 & b \\-\frac{1}{b} & 0 \end{array} \right) , \quad  \quad b = e^{{\rm sech}^{-1} \left( 2 \sin \frac{\pi}{q} \right) }  .
\ee
Clearly, $S^2 \cong 1$ in $PSL(2,\mathbb{R})$.   The action of $T$ is simply $T=U S$, and one can check that $T^3 = (US)^3 \cong 1$. 

Finally, we can translate the action of $S$ on the UHP to its action on the Poincar\'e disk:
\be
S: z \rightarrow \frac{z-\sqrt{2 \cos (2 \pi/q )-1}}{\sqrt{2 \cos (2 \pi/q)-1}\; z  -1} .
\ee 
As a check, note that $S$ takes the origin in $z$-coordinates to $z_1=\sqrt{2 \cos \left(\frac{2 \pi }{q}\right)-1}$.  By (\ref{eq:glob_metric}), the geodesic distance between the origin and $z_1$ is $2 \tanh^{-1} |z_1| = 2 \cosh^{-1} \frac{1}{2 \sin( \frac{\pi}{q})}$. This gives us an independent derivation for the length $a$ of the edges of the lattice, which agrees with  our previous formula (\ref{eq:latticeLength}) .
 In terms of this geodesic length $a$, the transformation $S$ 
can be  seen as a boost
\be
S: z \rightarrow \frac{\cosh (a/2) z - \sinh (a/2)  }{\sinh (a/2) z -  \cosh (a/2) } .
\ee

Because the lattice is an irreducible representation of $D(2,3,q)$, one can generate the entire lattice by starting with one point and repeatedly acting with $S$ and $T$.  So for instance, if we take our initial point to the the origin of the Poincar\'e disk, then acting with $S$ generates one vertex of the first  layer.  Acting with $T^n$ for $n=0, \dots, q-1$, this vertex is repeatedly rotated by $2\pi/q$ to fill out the rest of the first layer.  Then, we can act with $S$ again to generate points on the second layer, and acting multiple times with $T$ to fill out the second layer, and so on.

\section{Recursive Enumeration on the Triangulated Disk}
\label{app:Recursive}

The construction of our triangulated lattice on $\mathbb H^2$ is defined by recursively adding
one layer at time starting from  $r = 0$. Each layer has
$n(L)$ vertices on a periodic ring connected by single links
on the triangles between them.  The total number of links between two
layers  is $n(L) + n(L-1)$.  Consequently,  one obtains
a sum rule by counting the number of links (or  {\em flux})  through each layer.
Consider the number of vertices  $n(L-1)$ at layer $L-1$.
All the vertices must have exactly $q$ neighbors. Of the $q$ links (per vertex) to these neighbors, two are  around the
circumference of the layer, and $q-2$  
must either enter from layer $L-2$ or exit to layer $L$.
This gives the sum rule, 
\be
 (q-2)n(L-1)  =   [ n(L) + n(L-1) ] +  [ n(L-1) +
  n(L-2) ]  ,
\ee
or the two term recursion relation in (\ref{eq:Recursion}),
\be
 n(L) = (q - 4) n(L-1) - n(L-2) .
\ee
 The solution is the sum of two homogeneous 
powers,  $n(L) = c_+x_+^L + c_-x_-^L,$  where  
\be
x_\pm = 1/x_\mp = \frac{1}{2} \Big[(q - 4) \pm \sqrt{ (q-6)(q-2)}\Big]
\ee
are the roots of the quadratic equation $x^2 - (q-4)x +1 = 0$. To fix the coefficients $c_\pm$ we  need two starting
conditions. In our construction we  chose  the  q-fold  vertex at $r =
  0$, so the initial condition  is $n(0) = 0,\, n(1) = q$. So the particular the solution for our  D(2,3,7) lattice is
\be
 n(L) =  \frac{7}{\sqrt{5}} \Big[ (3/2 + \sqrt{5}/2)^L -  (3/2 - \sqrt{5}/2)^L \Big] .
\ee
However, if   we were
  to place a 3-fold circumcenter of the
 equilateral triangle at $r = 0$,  the initial condition would be 
$n(0) = 3, n(1) =  (q - 4)n(0)$.  An equivalent   approach  is to
build both initial conditions into a  generating function,
$N(z) = \sum^\infty _{i=0} n(i) z^i = ( n(0) + z (n(1) +(q -4)
  n(0)))/(( z  - x_+) (z - x_-))$,  computing  $n(L) = ( 2\pi i)^{-1}
  \oint dz z^{-1- L}N(z)$ by contour integration.

The total number of vertices in our D(2,3,q) disk is 
\be
V(q,L) = 1 +\sum^L_{i=1} n(i) .
\ee
By  Euler's identity for a triangulated disk, $V-E + F = 1$ (zero handles and one boundary)  the number of
edges (E) and faces (F) are also fixed. To include the
boundary term, it is convenient to first add a point vertex at
$\infty$,  starting with Euler's identity for a triangulated sphere obeying 
$V_{\circ} -E_{\circ} + F_{\circ} = 2 - H = 2$ and the constraint $2 E_{\circ}  =  3 F_{\circ}$. Then it is simple to solve these two equations for $E_0$ and $F_0$ in terms of $V_0$ to find $E_0 = 3 V_0 - 6 , F_0=2V_0 - 4$.  Deleting the links between the  $B(q,L) = n(L)$ boundary sites eliminates $B(q,L)$ edges and $B(q,L)$ faces, but only one vertex. Therefore the number of edges $E(q,L)$ and faces $F(q,L)$ in the disk with the point at infinity deleted are 
\be
 E(q,L) = 3V(q,L)  - 3 - B(q,L) ,  \quad   \quad   F(q,L) = 2 V(q, L)
 - 2 - B(q,L)  ,
 \label{eq:EFVqlSoln}
\ee
where $V(q,L)$ is the number of vertices in the graph. One can easily check that (\ref{eq:EFVqlSoln}) solves the
Euler identity on the disk.

Since all of these functions scale as $x^L_+$ for large $L$, we can obtain
rigorous scalings of our hyperbolic network toward the boundary. 
Assuming that the layers are on average spherical, a comparison is made
between the   area of the finite disk in the continuum, 
\be
A_{disk} = 2 \pi \int^{\rho_0}_0 d\rho \sinh(\rho)  = 2 \pi
(\cosh(\rho_0) -1)  ,
\ee
and the area of the triangulated lattice 
\be
A_{lat} =  A_\triangle(q)  F(q,L)  = (\pi
- 6 \pi /q)  F(q,L)  .
\ee
Matching the exponential growth for  $A_{disk}  \sim A_{lat}$ we find that
\be
\rho_0  \simeq \log(x_+) L  = \log[ (q- 4)/2 +\sqrt{
  (q-6)(q-2)}/2]  \; L,
\ee 
so that the effective lattice spacing between layers is
\be \label{eq:efflatspacing}
 a_q = \log[ (q- 4)/2 + \sqrt{ (q-6)(q-2)}/2]  .
\ee
For  $q = 7$  the effective  lattice spacing in $\rho$ is $ a_7
 =  0.962424$,  which is  consistent with the numerical estimate,
$\tilde c_7 = 0.919$,  from the finite L extrapolation in (\ref{eq:EpsScaling}).
 
With this identification of the lattice
spacing the solution for general $q$ is written, 
\be
n(L) = \frac{q}{2\sinh(a_q)} \Big[ e^{  a_q L} - e^{-a_q L} \Big] = \frac{ q  \sinh(a_q L)}{\sinh(a_q )}  .
\ee
The flat space limit, $q \rightarrow 6, a_q \rightarrow 0$, gives the correct enumeration, $n(L) = 6 L$, of vertices at each layer for the triangular lattice.

Several comments are worth making.  First it  is interesting to look at the $q$-dependence  for $a_q = 0.9624, \,
1.317,\, 1.567$    for  $q = 7,\, 8,\, 9$, respectively. Remarkably 
this is the same  lattice spacing $b =0.96,  \, 1.34,  \, 1.66$
determined  numerically in Sec.\ \ref{subsec:LargeMass} by fitting to the
Green's function for large AdS mass.

Next we note this recursive enumeration is  general for any
triangulated  planar graph. Starting from a node, each layer
is defined by single link to the next layer. Consequently this 
can be applied to the coarse graining of an ensemble
of  Regge calculus triangulations approximating a smooth manifold
with small average values of $\< q -6\> \simeq 0$. To this end, rewrite the recursion relation as
\be
n[L+1] - 2n[L] + n[L-1] = (q-6)n[L] ,
\ee
which approaches the following continuum equation, 
\be
 \partial^2_\rho n(\rho) \simeq  \frac{ (q-6)}{ a^2_q} n(\rho) \; ,
\ee
in the limit with  $a_q \simeq \sqrt{q-6}$ as $\<q \> \rightarrow 6$. So for  $\rho \gg a_q$ the 
continuum solution  $n(\rho)
 \sim e^{ \rho} $ gives precisely the right 
exponential growth  in arc length in $\theta$ for AdS$_2$  at fixed $\rho$.
Similar  methods  would also apply to our DEC  refinement scheme as it
approaches the continuum.

\bibliographystyle{utphys}
\bibliography{ads2lat}{}

\end{document}